\documentclass{article}

\usepackage{arxivsimplified}

\usepackage{moreverb}
\usepackage{cite}
\usepackage{amsmath,amsfonts,amssymb}
\usepackage{subfigure}
\usepackage{float}
\usepackage{color}
\usepackage{multirow}
\usepackage{graphicx}
\graphicspath{{Figures/}}
\usepackage{mathtools}
\usepackage[ruled,vlined]{algorithm2e}
\usepackage{enumitem}

\usepackage{booktabs}

\usepackage[colorlinks,bookmarksopen,bookmarksnumbered,citecolor=red,urlcolor=red]{hyperref}

\newcommand\BibTeX{{\rmfamily B\kern-.05em \textsc{i\kern-.025em b}\kern-.08em
T\kern-.1667em\lower.7ex\hbox{E}\kern-.125emX}}

\usepackage[utf8]{inputenc} 
\usepackage[T1]{fontenc}    
\usepackage{hyperref}       
\usepackage{url}            
\usepackage{booktabs}       
\usepackage{amsfonts}       
\usepackage{nicefrac}       
\usepackage{microtype}      
\usepackage{lipsum}

\usepackage{graphicx}
\usepackage{cite}
\usepackage[dvipsnames]{xcolor}
\usepackage{amsmath}
\usepackage{blindtext}
\usepackage{amsfonts}
\usepackage{multirow}
\usepackage{array}
\usepackage{mathtools}
\usepackage{listings}
\usepackage{xcolor}
\definecolor{codegreen}{rgb}{0,0.6,0}
\definecolor{codegray}{rgb}{0.5,0.5,0.5}
\definecolor{codepurple}{rgb}{0.58,0,0.82}
\definecolor{backcolour}{rgb}{0.95,0.95,0.92}

\lstdefinestyle{mystyle}{
    backgroundcolor=\color{backcolour},   
    commentstyle=\color{codegreen},
    keywordstyle=\color{magenta},
    numberstyle=\tiny\color{codegray},
    stringstyle=\color{codepurple},
    basicstyle=\ttfamily\footnotesize,
    breakatwhitespace=false,         
    breaklines=true,                 
    captionpos=b,                    
    keepspaces=true,                 
    numbers=left,                    
    numbersep=5pt,                  
    showspaces=false,                
    showstringspaces=false,
    showtabs=false,                  
    tabsize=2
}

\usepackage{scalerel,stackengine}
\stackMath
\newcommand\reallywidehat[1]{%
\savestack{\tmpbox}{\stretchto{%
  \scaleto{%
    \scalerel*[\widthof{\ensuremath{#1}}]{\kern-.6pt\bigwedge\kern-.6pt}%
    {\rule[-\textheight/2]{1ex}{\textheight}}
  }{\textheight}%
}{0.5ex}}%
\stackon[1pt]{#1}{\tmpbox}%
}

\title{Memory embedded non-intrusive reduced order modeling of non-ergodic flows}

\author{
  Shady~E.~Ahmed \\
  School of Mechanical \& Aerospace Engineering,\\
  Oklahoma State University, \\
  Stillwater, Oklahoma - 74078, USA.\\
  \texttt{shady.ahmed@okstate.edu} \\
  \And
 Sk. Mashfiqur Rahman \\
 School of Mechanical \& Aerospace Engineering,\\
  Oklahoma State University, \\
  Stillwater, Oklahoma - 74078, USA.\\
  \texttt{skmashfiqur.rahman@okstate.edu } \\
   \And
 Omer San \\
 School of Mechanical \& Aerospace Engineering,\\
  Oklahoma State University, \\
  Stillwater, Oklahoma - 74078, USA.\\
  \texttt{osan@okstate.edu} \\
  \And
  Adil Rasheed  \\
  Department of Engineering Cybernetics,\\
  Norwegian University of Science and Technology,\\
  N-7465, Trondheim, Norway.\\
  \texttt{adil.rasheed@ntnu.no }\\
  \And
  Ionel~M.~Navon \\
  Department of Scientific Computing,\\
  Florida State University, Tallahassee,\\
  Florida 32306 USA.\\
  \texttt{inavon@fsu.edu}\\
}

\begin{document}
\maketitle

\begin{abstract}
Generating a digital twin of any complex system requires modeling and computational approaches that are efficient, accurate, and modular. Traditional reduced order modeling techniques are targeted at only the first two but the novel non-intrusive approach presented in this study is an attempt at taking all three into account effectively compared to their traditional counterparts. Based on dimensionality reduction using proper orthogonal decomposition (POD), we introduce a long short-term memory (LSTM) neural network architecture together with a principal interval decomposition (PID) framework as an enabler to account for localized modal deformation, which is a key element in accurate reduced order modeling of convective flows. Our applications for convection dominated systems governed by Burgers, Navier-Stokes, and Boussinesq equations demonstrate that the proposed approach yields significantly more accurate predictions than the POD-Galerkin method, and could be a key enabler towards near real-time predictions of unsteady flows.       
\end{abstract}


\maketitle

\section{Introduction}
\label{sec:intro}

There are a great number of high-dimensional problems in the field of science and engineering (like atmospheric flows) that can be efficiently modeled based on embedded low-dimensional structures or reduced order models (ROMs) \cite{noack2003hierarchy,lucia2004reduced,quarteroni2014reduced}. Reduced order models have great promise for flow control\cite{noack2011reduced,ito2001reduced,mcnamara2004fluid,bergmann2008optimal,ravindran2000reduced,graham1999optimala,graham1999optimalb,brunton2015closed}, data assimilation \cite{daescu2007efficiency,navon2009data,cao2007reduced,he2011use,houtekamer1998data,houtekamer2001sequential,bennett2005inverse,evensen2009data,law2012evaluating,buljak2011inverse,xiao2018parameterised}, parameter estimation \cite{fu2018pod,boulakia2012reduced,kramer2017feedback}, and uncertainty quantification\cite{galbally2010non,biegler2011large,mathelin2005stochastic,smith2013uncertainty,najm2009uncertainty,zahr2019efficient}. These applications typically require multiple forward simulations of the problem being investigated and even the most powerful supercomputers might fail to perform these simulations in high-dimensional space, due to storage and speed limitations. Also, the building of digital twins \cite{glaessgen2012digital,grieves2017digital,boschert2016digital,uhlemann2017digital,tao2018digital} requires real-time and many-query responses. A digital twin can be defined as the virtual representation of a physical object or system across its lifecycle using real-time data \cite{grieves2011virtually} which requires efficient and on-the-fly simulation emulator. The concept of using such interactive computational megamodels often payoffs in terms of accelerated design cycle times, greater efficiency and safety, predictive maintenance and scheduling, more efficient and informed decision support systems, real-time monitoring, performance optimization, supervisory control to reduce energy consumption, and perhaps much beyond. With the recent wave of digitization, reduced order modeling can be viewed as one of the key enablers to bring the promise of the digital twinning concept closer to reality \cite{hartmann2018model}.  Therefore, there is a continuous demand for the development of accurate reduced order models for complex physical phenomena. In projection-based ROMs, the most widely used technique, the discrete high-dimensional operators are projected onto a lower-dimensional space, so that the problem can be solved more efficiently in this reduced space \cite{roychowdhury1999reduced,buffoni2006low,lakshmivarahan2008relation,fang2009pod,milk2016pymor,puzyrev2019pyrom}.



One of the very early-developed and well-known approaches to build this reduced space is Fourier analysis. However, it assumes universal basis functions (or modes) which have no specific relation to the physical system. On the other hand, snapshot-based model reduction techniques tailor a reduced space that best fits the problem by extracting the underlying coherent structures that controls the major dynamical evolution we are interested in. Proper orthogonal decomposition (POD) is a very popular and well-established approach extracting the modes which most contributes to the total variance \cite{volkwein2013proper,grassle2019model}. In fluid dynamics applications, where we are mostly interested in the velocity field, those modes contain the largest amount of kinetic energy \cite{taira2017modal,rowley2017model}. That is why POD is usually classified as an energy-based decomposition method. Another popular approach for model order reduction is the dynamic mode decomposition (DMD) \cite{schmid2010dynamic,schmid2011applications,chen2012variants,tu2013dynamic,kutz2016dynamic,bistrian2017method} which generates a number of modes, each characterized by an oscillating frequency and growth/decay rate. In the present study, we are interested in the application of POD for dimensionality reduction.

POD generates a set of spatial orthonormal basis functions, each containing a significant amount of total energy. To obtain a reduced representation of a system, the first few modes are selected, and the remaining are truncated assuming their contribution to the system's behavior is minimum (i.e., compression). The Kolmogorov $n$-width \cite{kolmogoroff1936uber,pinkus2012n} provides a mathematical guideline to quantify the optimal n-dimensional linear subspace and the associated error (i.e., a measure of system's reducibility).
It is a classical concept of approximation theory which describes the error (in worst-case scenario) that might arise from a projection onto the best-possible subspace of a given dimension $n$ \cite{greif2019decay,pinkus2012n}. If the decay of Kolmogorov width is fast, then employing a reduced linear subspace is feasible. Unfortunately, this is not generally the case for convection-dominated flows with severe temporal evolution, or equivalently parametric situations where the solution is highly dependent on the parameter space. In these situations, the decay of $n$-width is relatively slow, hence raising the Kolmogorov barrier (i.e., requiring more modes to be retained in the reduced-space approximation of the underlying dynamics). Instead of working with linear manifolds, Lee and Carlberg \cite{lee2018model} proposed a projection onto nonlinear manifolds to break this barrier. 

Moreover, snapshots-based model reduction techniques rely, in principle, on the ergodicity hypothesis which implies that any collection of random samples should be able to represent the whole process. That is the system's response given certain inputs is considered to encapsulate the essential behavior and characteristics of that system. The flow situations described before are usually non-ergodic and do not fulfill this hypothesis. As a consequence, the resulting intrinsically global POD modes cannot describe the underlying flow structures. In problems with strong convective nature, the system's condition is significantly different at different time instances. In addition, a global POD application on these systems causes averaging and deformation of POD modes in such a way that they become no more representative of any of the system's states. Principal interval decomposition (PID) \cite{borggaard2007interval,san2015principal,ahmed2018stabilized} offers a treatment of this modal deformation by dividing the temporal space into a few intervals, where POD (or any other decomposition technique) is applied locally.
Therefore, the local POD modes are tailored to the specific flow behavior in the respective sub-intervals. Similarly, local POD modes can be constructed by partitioning the physical domain \cite{taddei2015reduced,xiao2019domain}, state space \cite{amsallem2012nonlinear,peherstorfer2015online}, or parameter space. Indeed, this partitioning approach helps to break the Kolmogorov barrier as well. Using a single fixed global subspace would necessitate keeping a larger number of modes to meet accuracy requirements and capture a certain amount of energy. Alternatively, partitioning allows the use of several tailored, local, and lower-dimensional subspaces.

POD has been often coupled with Galerkin projection to build ROMs for linear and nonlinear systems \cite{ito1998reduced,iollo2000stability,rowley2004model,pinnau2008model, stankiewicz2008reduced,sachs2010pod,akhtar2009stability,barone2009stable}. In Galerkin projection, the governing equations are projected onto the POD subspace. Through the orthonormality and energy-optimality characteristic of POD modes, a simpler and truncated set of coupled ordinary differential equations (ODEs) is obtained. The resulting system is low-dimensional, but it is dense. In other words, it generates small and full matrices, while the common discretization techniques often lead to large and sparse matrices. Also, the quadratic non-linearity and triadic interactions in ROMs leads to a computational load of an order of $O(R^3)$, where $R$ is the number of retained modes. The discrete empirical interpolation method \cite{chaturantabut2010nonlinear} (DEIM) can be used to reduce the computational complexity for nonlinear ROM where the nonlinear term is approximated with sparse sampling through projecting it onto a separate reduced subspace, rather than directly computing it\cite{cstefuanescu2013pod,antil2014application,stanko2016nonlinear}. Also, symbolic regression techniques might serve to identify ROMs from limited sensor data \cite{loiseau2018sparse}. Another class for ROM which is gaining popularity in recent years, is the fully data-driven, non-intrusive ROM (NIROM) \cite{xie2019non,xiao2017towards,lin2017non,xiao2017parameterized,xiao2017non,xiao2016non,xiao2015nonRBF,xiao2015non,peherstorfer2019sampling}. NIROM is a family of methods that solely access available datasets to extract and mimic system's dynamics, with little-to-no knowledge of the governing equations. Non-intrusive approach is sometimes called physics-agnostic modeling, in contrast to the intrusive physics-informed approach. 

One of the main advantages of a non-intrusive approach is its portability, which results from the fact that it does not necessarily require the exact form of the equations and the methods used to solve them to generate the snapshots. This makes the approach applicable to experimental data where the equations are often not well established or have huge uncertainties involved in their parameters. Together with their modularity and simplicity, non-intrusive models offer a unique advantage in multidisciplinary collaborative environments. It is often necessary to share the data or the model without revealing the proprietary or sensitive information. Different departments or subcontractors can easily exchange data (with standardized I/O) or executables, securing their intangible assets and intellectual property rights. Furthermore, non-intrusive approaches are particularly useful when the detailed governing equations of the problem are unknown. This modeling approach can benefit from the enormous amount of data collected from experiments, sensor measurements, and large-scale simulations to build a robust and accurate ROM technology. 

With the growing advancement of artificial neural networks (ANNs) and other machine learning (ML) techniques, and the availability of massive amounts of data resources from high-fidelity simulations, field measurements, and experiments, the data-driven, non-intrusive modeling approaches are currently considered some of the most promising methods across different scientific and research communities. In the past few years, there have been a significant amount of research using ANN and ML techniques dedicated to turbulence modeling \cite{raissi2019physics,maulik2017neural,raissi2018multistep,lee1997application,maulik2018data,faller1997unsteady,san2018neural,wang2019non,moosavi2015efficient,kani2017dr}. More details on the influence of ML on fluid mechanics, specifically turbulence modeling can be found elsewhere\cite{brunton2019machine,kutz2017deep,durbin2018some,duraisamy2019turbulence,gamboa2017deep}. 
Until recently, the fully non-intrusive modeling can be considered most attractive enabling methodology to do real-time simulation very efficiently in the context of emerging digital twin technologies\cite{pawar2019deep}. 
In a complimentary fashion, the hybrid models\cite{krasnopolsky2006complex,krasnopolsky2006new,rahman2018hybrid,wan2018data,xie2018data,pathak2018hybrid} are developed by combining the intrusive and non-intrusive models in such way that the limitation of one component modeling strategy can be addressed by the other component model. 

In this work, we propose a non-intrusive reduced order modeling framework that is best suited for unsteady flows, where the convective mechanisms are more predominant than the diffusive ones. The approach is based on principal interval decomposition to parse the data over time to learn more localized dominant structures. This is particularly important for problems where we observe relatively slow decay in the Kolmogorov $n$-width, which constraints the feasibility of reduced-order approximation. Also, this partitioning helps to satisfy the ergodicity hypothesis within each local interval to provide a good approximation of the flow field. We couple this parsing technique with long short-term memory (LSTM) neural network, which is a very efficient ML tool for time-series predictions. The PID-LSTM is compared with its counterpart based on Galerkin projection (PID-GP). Not only PID-LSTM eliminates the need to access the governing equations, being solely dependent on data, but also it helps mitigate instabilities in the ROM predictions resulting from the non-linear interaction among different fields. We tested the proposed framework with one-dimensional and two-dimensional convective-dominated problems, highlighting its benefits over the standard POD and Galerkin projection approaches.

The rest of the paper is outlined here. Section~\ref{sec:mat} provides a brief overview of the mathematical models used to test the proposed framework. Namely, we describe the one-dimensional nonlinear advective Burgers problem, a standard benchmark problem in CFD studies. Also, we tested our framework using the two-dimensional Navier-Stokes equations. More specifically, we investigate the vortex merger and double shear layer problems. As a final and more complicated problem, we study the unsteady lock-exchange flow problem (also known as the Marsigli flow) governed by the two-dimensional Boussinesq equations. In Section~\ref{sec:pod}, we describe the classical proper orthogonal decomposition approach for order reduction. A generalized PID framework, an approach to construct local basis rather than global ones, is shown in Section~\ref{sec:pid}. The application of PID with classical Galerkin projection to build intrusive ROMs is outlined in Section~\ref{sec:GP}. The proposed approach which incorporates PID while bypassing GP in a non-intrusive framework is illustrated in Section~\ref{sec:LSTM}. Results obtained with the novel proposed method to illustrate its advantages are followed by relevant discussions in Section~\ref{sec:results} with relevant discussions. Finally, conclusions are provided in Section~\ref{sec:conc}.

\section{Mathematical models} \label{sec:mat}
\subsection{1D Burgers equation} \label{sec:burg}
Our first test case is the one-dimensional viscous Burgers equation. It represents a simple form of Navier-Stokes equations in a 1D setting with similar quadratic nonlinear interaction and Laplacian dissipation. It is therefore considered as a standard benchmark for the analysis of nonlinear advection-diffusion problems.

The evolution of the velocity field $u(x, t)$, in a dimensionless form, is given by
\begin{equation}
    \dfrac{\partial u}{\partial t} + u \dfrac{\partial u}{\partial x} = \dfrac{1}{\text{Re}} \dfrac{\partial ^2 u}{\partial x^2}, \label{eq:brg}
\end{equation}
where $\text{Re}$ is the dimensionless Reynolds number, defined as the ratio of inertial effects to viscous effects.

\subsection{2D Navier-Stokes equations} \label{sec:NS}
The primitive formulation of the 2D Navier-Stokes equations, in dimensionless form with index notation, can be written as
\begin{align}
    \dfrac{\partial u_i}{\partial x_i} &= 0,  \\
    \dfrac{\partial u_i}{\partial t} + u_j\dfrac{\partial u_i}{\partial x_j} &= -\dfrac{\partial p}{\partial x_i} + \dfrac{1}{\text{Re}}\dfrac{\partial^2 u_i}{\partial x_j\partial x_j } \label{eq:NSmom}, 
\end{align}
where $u_i$ is the velocity in the $i$-th direction, and $p$ is the pressure. Alternatively, by taking the curl of Equation~\ref{eq:NSmom}, the following vorticity-streamfunction formulation of the 2D Navier-Stokes equation is obtained,
\begin{align} \label{eq:NS}
\dfrac{\partial \omega}{\partial t} + \dfrac{\partial \psi}{\partial y}\dfrac{\partial \omega}{\partial x} - \dfrac{\partial \psi}{\partial x} \dfrac{\partial \omega}{\partial y} &= \dfrac{1}{\text{Re}}\left( \dfrac{\partial^2 \omega}{\partial x^2} + \dfrac{\partial^2 \omega}{\partial y^2} \right) ,
\end{align}
where $\omega$ is the vorticity defined as $\omega = \nabla \times \mathbf{u}$, $\mathbf{u} = [u,v]^T$ is the velocity vector, and $\psi$ is the streamfunction. The vorticity-streamfunction formulation has several computational advantages over the primitive variable formulation. It prevents the odd-even decoupling issues that might arise between pressure and velocity components. Therefore, a collocated grid can be used instead of using a staggered one without producing any spurious modes. Also, it automatically enforces the incompressibility condition. The kinematic relationship between vorticity and streamfunction is given by the following Poisson equation,
\begin{equation}\label{eq:Poisson}
\dfrac{\partial^2 \psi}{\partial x^2} + \dfrac{\partial^2 \psi}{\partial y^2} = -\omega.
\end{equation}
This relationship implies that the streamfunction is not a prognostic variable, and can be computed from the vorticity field at each timestep. We will also use this property in our development of the intrusive ROMs with Galerkin projection in Section~\ref{sec:GP}. If we define the Jacobian, $J(f,g)$ and the Laplacian $\nabla^2 f$ operators as follows
\begin{align}
    J(f,g) &= \dfrac{\partial f}{\partial x} \dfrac{\partial g}{\partial y} -  \dfrac{\partial f}{\partial y} \dfrac{\partial g}{\partial x}, \\
    \nabla^2 f &= \dfrac{\partial^2 f}{\partial x^2} + \dfrac{\partial^2 f}{\partial y^2},
\end{align}
Equations~\ref{eq:NS} and~\ref{eq:Poisson} can be rewritten as
\begin{align} 
\dfrac{\partial \omega}{\partial t} + J(\omega,\psi) &= \dfrac{1}{\text{Re}} \nabla^2 \omega, \label{eq:NS2} \\
\nabla^2 \psi &= -\omega.
\end{align}

\subsection{2D Boussinesq equations} \label{sec:Bouss}
Boussinesq equations represent a simple approach for modeling geophysical waves such as oceanic and atmospheric circulations induced by temperature differences \cite{majda2003introduction}. The Boussinesq approximation enables us to solve non-isothermal flows (e.g., natural convection), without having to solve for the full compressible formulation of Navier-Stokes equations. In this approximation, variations of all fluid properties other than the density are ignored completely. Moreover, the density dependence is ignored in all terms except for gravitational force (giving rise to buoyancy effects). As a result, the continuity equation is used in its constant density form, and the momentum equation can be simplified significantly.

The dimensionless form of the two-dimensional incompressible Boussinesq equations on a domain $\Omega$ in vorticity-streamfunction formulation is given by the following the two coupled scalar transport equations \cite{liu2003fourth,nicolas20052d},
\begin{align}
\dfrac{\partial \omega}{\partial t} + J(\omega,\psi) &= \dfrac{1}{\text{Re}}\nabla^2 \omega + \text{Ri} \dfrac{\partial \theta}{\partial x}, \label{eq:Bouss1} \\
\dfrac{\partial \theta}{\partial t} + J(\theta,\psi) &= \dfrac{1}{\text{Re} \text{Pr}} \nabla^2 \theta, \label{eq:Bouss2}
\end{align}
where $\theta$ is the temperature. In Boussinesq flow systems, there are  three leading physical mechanisms, namely viscosity, conductivity, and buoyancy. Equations~\ref{eq:Bouss1}-\ref{eq:Bouss2} include three dimensionless numbers; $\text{Re}$, $\text{Ri}$, and $\text{Pr}$. $\text{Re}$ is the dimensionless Reynolds number, relating viscous effects and inertial effects as defined in Section~\ref{sec:burg}. Richardson number, $\text{Ri}$ is the ratio of buoyancy force to inertial force and Prandtl number, $\text{Pr}$, is the ratio between kinematic viscosity and heat conductivity. Boussinesq approximation underlies the statement that dynamical similarity of free convective flows depends on the Grashof and Prandtl numbers \cite{tritton1988physical}, where Grashof number, $\text{Gr}$, is defined as $\text{Gr} = \text{Ri} \text{Re}^2$. In natural convection heat transfer, other relevant dimensionless numbers are used, such as Rayleigh number, $\text{Ra} = \text{Gr} \text{Pr}$, and P\'eclet number, $\text{Pe} = \text{Re} \text{Pr}$.

\section{Proper orthogonal decomposition} \label{sec:pod}
In POD, the dominant spatial subspaces are extracted from a given dataset. In other words, POD computes the dominant coherent directions in an infinite space which best describe the spatial evolution of a system. POD-ROM is closely-related to either singular value decomposition or eigenvalue decomposition of snapshot matrix (in finite-dimensional case). However, in most fluid flow simulations of interest, the number of degrees of freedom (number of grid points) is often orders of magnitude larger than number of collected datasets. This results in a tall and skinny matrix, which makes the conventional direct decomposition inefficient, as well as time and memory consuming. Therefore, we follow the method of snapshots, proposed by Sirovich \cite{sirovich1987turbulence}, to generate the POD basis efficiently. A number of snapshots (or realizations), $N_s$, of the flow field, denoted as $u(\mathbf{x},t_n)$, are stored at consecutive times $t_n$ for $n=0,1,2,\dots,N_s$. The field $u(\mathbf{x},t_n)$ is assumed to be square-integrable in the Hilbert space, $u(\mathbf{x},t_n) \in L^2(\cal{D},\cal{T})$ with $\mathbf{x} \in {\cal{R}}^N, N=1, 2, 3, \dots$ and $t_n\in {\cal{T}} = [0,T]$. From physical point of view, square-integrability corresponds to a finite amount of kinetic energy in the field. The time-averaged field, called `base flow', can be computed as
\begin{equation}
\bar{u}(\mathbf{x})=\dfrac{1}{N_s}\sum_{n=0}^{N_s}u(\mathbf{x}, t_n).
\end{equation}
The mean-subtracted snapshots, also called anomaly or fluctuation fields, are then computed as the difference between the instantaneous field and the mean field
\begin{equation}
u'(\mathbf{x}, t_n)=u(\mathbf{x}, t_n)-\bar{u}(\mathbf{x}).
\end{equation}
This subtraction has been common in ROM community, and it guarantees that ROM solution would satisfy the same boundary conditions as the full order model \cite{chen2012variants}. This anomaly field procedure can be also interpreted as a mapping of snapshot data to its origin.

Then, an $N_s\times N_s$ snapshot data matrix $\mathbf{A}=[a_{ij}]$  is computed from the inner product of mean-subtracted snapshots
\begin{equation}
a_{ij}=\langle u'(\mathbf{x}, t_i); u'(\mathbf{x}, t_j)\rangle,
\end{equation}
where the angle-parenthesis denotes the inner product defined as 
\begin{equation}
    \langle q_1(\mathbf{x}); q_2(\mathbf{x})\rangle = \int_{\Omega}{q_1(\mathbf{x})q_2(\mathbf{x}) d\mathbf{x}}.
\end{equation}
It turns out that $\mathbf{A}$ is a non-negative, postive-semidefinite Hermitian matrix. An eigenvalue decomposition of $\mathbf{A}$ is carried out as,
\begin{equation}
\mathbf{A} \mathbf{V} = \mathbf{V} \mathbf{\Lambda},
\end{equation} 
where $\mathbf{\Lambda}$ is a diagonal matrix whose entries are non-negative eigenvalues $\lambda_k$ of $\mathbf{A}$, and $\mathbf{V}$ is a matrix whose columns $\mathbf{v}_k$ are the corresponding eigenvectors. It should be noted that these eigenvalues need to be arranged in a descending order (i.e., $\lambda_1\ge\lambda_2\ge\dots\ge\lambda_{N_s}$), for proper selection of the POD modes. In general, the eigenvalues, $\lambda_k$, represent the respective POD mode contribution to the total variance. In case of velocity time series, it represents the turbulent kinetic energy. The POD modes $\phi_{k}$ are then computed as
\begin{equation}
\phi_{k}(\mathbf{x})=\dfrac{1}{\sqrt{\lambda_k}}\sum_{n=1}^{N_s} v^{n}_{k} u'(\mathbf{x}, t_n),
\end{equation}
where $v^{n}_{k}$ is the $n$-th component of the eigenvector $\mathbf{v}_k$. The scaling factor, $1/\sqrt{\lambda_k}$, is to guarantee the orthonormality of POD modes, i.e., $\langle \phi_i; \phi_j\rangle = \delta_{ij}$, where $\delta_{ij}$ is the Kronecker delta. Using this basis, we can represent our reduced-order approximation of the field as follows,
\begin{align} \label{eq:upod}
u(\mathbf{x}, t_n) &= \bar{u}(\mathbf{x}) + \sum_{k=1}^{R} \alpha_k(t_n) \phi_k(\mathbf{x}),\\
\alpha_k(t_n) &=  \big \langle u(\mathbf{x}, t_n) - \bar{u}(\mathbf{x}) ;  \phi_k(\mathbf{x}) \big \rangle,
\end{align}
where $R$ is the number of retained modes ($R<< N_s << N$), where $N_s$ is the number of collected snapshots and $N$ is the spatial dimension (number of grid points). POD is optimal in the sense that it provides the most efficient way (with respect to other linear representations) of capturing the dominant components of an infinite-dimensional process with only finite, and few, modes. The POD modes can be interpreted geometrically as the principal axes of the cloud of data points, $\{u(\mathbf{x}, t_n)\}_{n=1}^{N_s}$ in the $N$-dimensional vector space.

From a mathematical perspective, the set of POD modes, $\mathbf{\Phi} =\{\phi_k\}_{k=1}^{R}$, represents the solution of the following optimization problem \cite{volkwein2013proper},
\begin{align*}
    \max \quad & \sum_{n=0}^{N_s} \sum_{k=1}^{R} \bigg| \big\langle u'(\mathbf{x}, t_n); \phi_k(\mathbf{x})\big\rangle \bigg|^2,\\
    \text{subject to} \quad& \phi_k \in L^2({\cal{D}}), \quad \big \langle \phi_i ; \phi_j \big \rangle = \delta_{ij}.
\end{align*}
A more in-depth discussion about mathematical aspects of POD and its optimality can be found in the rigorous discussions by Berkooz et al \cite{berkooz1993proper} and Holmes et al \cite{holmes2012turbulence}. 

\section{Principal interval decomposition} \label{sec:pid}
The classical POD approach, presented in Section~\ref{sec:pod}, produces a set of modes, or basis, that contains the largest amount of snapshot energy, but in average sense. Intrinsically, this results in global modes that are most similar to the overall flow. In general, POD-based ROMs work well for relatively smooth and ergodic systems with rapid decay in the Kolmogorov $n$-width. For those, only the first few $R$ modes are sufficient to represent the system with acceptable accuracy, where $R<<N_s<<N$, and the remaining $(N_s-R)$ modes are truncated with minimal effect. However, in nonlinear convective flow problems, this is not always the case. As a consequence, energy is widely distributed over a large number of modes. Therefore, the truncated modes possess a significant amount of total energy and an increased number of modes need to be retained in order to describe the system in hand properly. Closure models and stabilization schemes were proven to improve the performance of ROMs via approximating the effects of truncated modes \cite{kalb2007intrinsic,bergmann2009enablers,borggaard2011artificial,wang2012proper,rahman2019dynamic}. Despite this, the accuracy gain from closure and stabilization only is limited in these scenarios. 

A major source of accuracy loss in the application of POD in convection-dominated flows comes from the global nature of POD approach, which results in overall deformation of the obtained modes by the rapidly varying flow field state. As a result, the constructed POD modes do not resemble any of the flow states and cannot capture any dominant structure at all. Furthermore, excursions in state space that contain a small amount of energy can be overlooked by POD because their contribution to the total energy may be negligible. These excursions can, however, be of interest and have significant impact on the dynamical evolution (see Cazemier et al \cite{cazemier1998proper} for example). To address these issues, we follow and extend the principal interval decomposition (PID) approach, first presented by IJzerman \cite{ijzerman2000signal}. The main purpose of PID is to replace the global POD modes, with localized ones. This is accomplished by dividing the whole time domain ${\cal{T}}$ into a number $N_p$ of non-overlapping time windows, $\tau_1, \tau_2, \dots, \tau_{N_p}$ where ${\cal{T}} = \cup_{p=1}^{N_p} \tau_p$. We denote $\kappa^{(p)}$ as the index of the time instance at the interface between the consecutive sub-intervals $\tau_p$ and $\tau_{p+1}$ (i.e., $\tau_p \cap \tau_{p+1} = t_{\kappa^{(p)}}$, $p=1,2,\dots,N_p-1$).  That is,
\begin{equation}
    \tau_p = [t_{\kappa^{(p-1)}},t_{\kappa^{(p)}}],
\end{equation}
Here, we restrict ourselves to equally space time intervals although an adaptive partitioning approach may be performed \cite{dihlmann2011model}. A set of local basis functions $\mathbf{\Phi}^{(p)}=\{\phi^{(p)}_k\}_{k=1}^{R^{(p)}}$ is constructed for each time window $\tau_p$, following the same standard procedure described in Section~\ref{sec:pod} within each sub-interval, where $\phi_k^{(p)}$ is the $k$-th mode in the $p$-th interval and $R^{(p)}$ is the number of modes in this interval. Even though the approach is applicable for different numbers of modes in each interval, we will continue our discussion assuming fixed number of modes per interval (i.e., $R^{(p)} = R$, $p=1,2,\dots,N_p$). It should be noted that local mean fields are also constructed within each interval as,
\begin{equation}
\bar{u}^{(p)}(\mathbf{x})=\dfrac{1}{N_s/N_p}\sum_{n=\kappa^{(p-1)}}^{\kappa^{(p)}}u(\mathbf{x}, t_n),
\end{equation}
where $\bar{u}^{(p)}(\mathbf{x})$ is the mean field over the sub-interval $\tau_p$. Our reduced-rank approximation becomes
\begin{align} \label{eq:upid}
u(\mathbf{x}, t_n) &= \bar{u}^{(p)}(\mathbf{x}) + \sum_{k=1}^{R} \alpha^{(p)}_k(t_n) \phi^{(p)}_k(\mathbf{x}), \\
\alpha^{(p)}_k(t_n) &= \big\langle u(\mathbf{x}, t_n) - \bar{u}^{(p)}(\mathbf{x}) \label{eq:upid2}  ; \phi^{(p)}_k(\mathbf{x}) \big\rangle, \nonumber \\
& \quad \forall \ t_{\kappa^{(p-1)}} \le t_n \le t_{\kappa^{(p)}}.
\end{align}
Although it might seem that PID would be more computationally costly (implementing the standard POD procedure $N_p$ times), the actual time to perform PID is reduced, as reported in Section~\ref{sec:results}. This is caused by solving a number $N_p$ of smaller ($N_s/N_p \times N_s/N_p$) eigenvalue problems rather than solving one big ($N_s \times N_s$) eigenvalue problem \cite{san2015principal}.

\section{Intrusive reduced order modeling}
\label{sec:GP}
In order to build intrusive reduced order models, we apply standard Galerkin projection to our nonlinear systems presented in Section~\ref{sec:mat}. First the governing equations (i.e., Equations~\ref{eq:brg},~\ref{eq:NS2},~\ref{eq:Bouss1}, and~\ref{eq:Bouss2}) need to be rearranged in semi-discretized form using linear and nonlinear operators as follows,
\begin{align} \label{eq:oper}
    \dfrac{\partial {q}}{\partial t} &= {{\cal{L}}} + {{\cal{N}}},
\end{align}
where $q$ is $u$ in Burgers equation and $\omega$ in Navier-Stokes problem, and $[\omega, \theta]$ in Boussinesq case. The linear and non-linear operators are summarized in Table~\ref{tab:operators}. Then, the reduced-order approximation (i.e., Equation~\ref{eq:upod} or~\ref{eq:upid}) is plugged into Equation~\ref{eq:oper} and Galerkin projection is applied by multiplying Equation~\ref{eq:oper} with the basis functions $\phi^{(p)}_k$ and integrating over the domain. The orthogonrmality property of POD modes can be used to reduced the equations into a set of coupled ordinary differential equations (ODEs) in the POD coefficients, $\alpha_k$.

\begin{table}[ht!]
	\caption{Linear and nonlinear operators for mathematical models introduced in Section~\ref{sec:mat}.}
	\centering
	\begin{tabular}[t]{l l  l } 
		\hline\noalign{\smallskip}
		${q}$ \qquad \qquad \qquad \qquad & \quad ${{\cal{L}}}$ \qquad \qquad \qquad \qquad \qquad & \quad ${{\cal{N}}}$ \qquad \quad \\ [0.8ex] 
		\hline\noalign{\smallskip}
		\multicolumn{3}{l}{\emph{\underline{1D Burgers}}} \smallskip \\
		$u$ & $\dfrac{1}{\text{Re}} \dfrac{\partial^2 u}{\partial x^2}$ & $ -u \dfrac{\partial u}{\partial x}$   \bigskip \\ 
		\multicolumn{3}{l}{\emph{\underline{2D Navier-Stokes}}} \smallskip \\
		$\omega$ & $ \dfrac{1}{\text{Re}} \nabla^2 \omega $ & $-J(\omega,\psi)$      \bigskip \\ 
		\multicolumn{3}{l}{\emph{\underline{2D Boussinesq}}} \smallskip \\
		$\omega$ & $ \dfrac{1}{\text{Re}} \nabla^2 \omega + \text{Ri} \dfrac{\partial \theta}{\partial x}$ & $-J(\omega,\psi)$  \medskip \smallskip  \\ 
		$\theta$ & $\dfrac{1}{\text{Re} \text{Pr} } \nabla^2 \theta$   & $-J(\theta,\psi)$  \medskip \\ 
		\hline
	\end{tabular}
	\label{tab:operators}
\end{table}

A summary of the obtained ROM equations from applying principal interval decomposition approach with Galerkin projection (PID-GP), is given below. Details of the derivation can be found elsewhere \cite{san2013proper, rahman2019dynamic,san2015principal}. 

\subsection{1D Burgers equation}
The reduced-rank approximation for Burgers problem as long as the dynamical evolution equation for the temporal coefficients using PID approach coupled with Galerkin projection (PID-GP) can be written as,
    \begin{align}
        u(x,t) &= \bar{u}^{(p)}(x) + \sum_{k=1}^R \alpha^{(p)}_k(t) \phi^{(p)}_k(x) \nonumber \\
        & \qquad \forall \	t_{\kappa(p-1)} \le t \le t_{\kappa(p)}, \\
        \dfrac{\text{d}\alpha^{(p)}_k}{\text{d}t} &=  \mathfrak{B}^{(p)}_k + \sum_{i=1}^{R} \mathfrak{L}^{(p)}_{i,k} \alpha^{(p)}_i + \sum_{i=1}^{R} \sum_{j=1}^{R} \mathfrak{N}^{(p)}_{i,j,k} \alpha^{(p)}_i \alpha^{(p)}_j, \nonumber \\
        & \qquad k=1,2,\dots,R,\label{eq:rombrg}
    \end{align}
    where $\mathfrak{B}$, $\mathfrak{L}$, and $\mathfrak{N}$ are predetermined model coefficients corresponding to constant, linear and nonlinear terms, respectively. They are precomputed only once during offline training phase as follows,
    \begin{align*}
        \mathfrak{B}^{(p)}_k &= \big\langle  \dfrac{1}{\text{Re}} \dfrac{\partial^2 \bar{u}^{(p)} }{\partial x^2} - \bar{u}^{(p)} \dfrac{\partial \bar{u}^{(p)}}{\partial x}; \phi^{(p)}_k \big\rangle, \\
        \mathfrak{L}^{(p)}_{i,k} & = \big\langle \dfrac{1}{\text{Re}} \dfrac{\partial^2 \phi_i^{(p)} }{\partial x^2} - \bar{u}^{(p)} \dfrac{\partial \phi_i^{(p)}}{\partial x} - \phi_i^{(p)} \dfrac{\partial \bar{u}^{(p)}}{\partial x};  \phi^{(p)}_k \big\rangle, \\
        \mathfrak{N}^{(p)}_{i,j,k} &= \big\langle -\phi_i^{(p)} \dfrac{\partial \phi_j^{(p)}}{\partial x};  \phi^{(p)}_k \big\rangle.
    \end{align*}
    For sake of brevity in 2D cases, we shall drop the superscript $(p)$ in the ROM equations below, but it should be noted that they are applicable interval-wise similar to the equations above.
\subsection{2D Navier-Stokes equations} 
In 2D Navier-Stokes equations, similar reduced-rank approximation and temporal evolution can be written as follows (after dropping the superscript $(p)$),
    \begin{align}
        \omega(x,y,t) &= \bar{\omega}(x,y) + \sum_{k=1}^R \alpha_k(t) \phi_k^{\omega}(x,y), \\
        \psi(x,y,t_n) &= \bar{\psi}(x,y) + \sum_{k=1}^R \alpha_k(t) \phi_k^{\psi}(x,y), \\
        \dfrac{\text{d}\alpha_k}{\text{d}t} &=  \mathfrak{B}_k + \sum_{i=1}^R \mathfrak{L}_{i,k} \alpha_i + \sum_{i=1}^R \sum_{j=1}^R \mathfrak{N}_{i,j,k} \alpha_i \alpha_j,
    \end{align}
    where
    \begin{align*}
        \mathfrak{B}_k &= \big\langle -J(\bar{\omega},\bar{\psi}) + \dfrac{1}{\text{Re}} \nabla^2 \bar{\omega} ; \phi_k^{\omega} \big\rangle, \\
        \mathfrak{L}_{i,k} &= \big\langle -J(\bar{\omega},\phi_i^{\psi})  -J(\phi_i^{\omega},\bar{\psi}) + \dfrac{1}{\text{Re}} \nabla^2 \phi_i^{\omega} ; \phi_k^{\omega} \big\rangle, \\
        \mathfrak{N}_{i,j,k} &= \big\langle -J(\phi_i^{\omega},\phi_j^{\psi}) ; \phi_k^{\omega} \big\rangle.
    \end{align*}
    We can observe that the vorticity and streamfunction share the same time-dependent coefficients because they are related through a kinematic relationship, given by Equation~\ref{eq:Poisson}. Moreover the mean field and spatial POD modes for streamfunction can be obtained from solving the following Poisson equations during offline stage because POD preserves linear properties,
    \begin{align}
        &\nabla^2 \bar{\psi}(x,y) = -\bar{\omega}(x,y), \\
        &\nabla^2 \phi_k^{\psi}(x,y) = -\phi_k^{\omega}(x,y), \quad k=1,2,\dots, R.
    \end{align}
    This result in a set of basis functions for the streamfunction that are not necessarily orthonormal. The same procedure will be used in 2D Boussinesq problem since it is also represented in vorticity-streamfunction formulation.
\subsection{2D Boussinesq equations} 
For 2D Boussinesq equations, the vorticity, streamfunction, and temperature fields can be written as,
    \begin{align}
        \omega(x,y,t) &= \bar{\omega}(x,y) + \sum_{k=1}^R \alpha_k(t) \phi_k^{\omega}(x,y), \\
        \psi(x,y,t_n) &= \bar{\psi}(x,y) + \sum_{k=1}^R \alpha_k(t) \phi_k^{\psi}(x,y), \\
        \theta(x,y,t_n) &= \bar{\theta}(x,y) + \sum_{k=1}^R \beta_k(t) \phi_k^{\theta}(x,y),
    \end{align}
    where the temporal coefficients $\alpha_k(t)$ and $\beta_k(t)$ can be calculated from the following ODEs,
    \begin{align}
        \dfrac{\text{d}\alpha_k}{\text{d}t} &=  \mathfrak{B}^{(\omega)}_k + \sum_{i=1}^R \mathfrak{L}^{(\omega,\psi)}_{i,k} \alpha_i + \sum_{i=1}^R \mathfrak{L}^{(\omega,\theta)}_{i,k} \beta_i \nonumber \\
        &+ \sum_{i=1}^R \sum_{j=1}^R \mathfrak{N}^{(\omega,\psi)}_{i,j,k} \alpha_i \alpha_j, \\
        \dfrac{\text{d}\beta_k}{\text{d}t} &=  \mathfrak{B}^{(\theta)}_k + \sum_{i=1}^R \mathfrak{L}^{(\theta,\psi)}_{i,k} \alpha_i + \sum_{i=1}^R \mathfrak{L}^{(\psi,\theta)}_{i,k} \beta_i \nonumber \\
        &+ \sum_{i=1}^R \sum_{j=1}^R \mathfrak{N}^{(\theta,\psi)}_{i,j,k} \alpha_i \beta_j,
    \end{align}
        where the predetermined coefficients are calculated as
    \begin{align*}
        \mathfrak{B}^{(\omega)}_k &= \big\langle -J(\bar{\omega},\bar{\psi}) + \dfrac{1}{\text{Re}} \nabla^2 \bar{\omega} + \text{Ri} \dfrac{\partial \bar{\theta}}{\partial x}; \phi_k^{\omega} \big\rangle, \\
        \mathfrak{B}^{(\theta)}_k &= \big\langle -J(\bar{\theta},\bar{\psi}) + \dfrac{1}{\text{Re}\text{Pr}} \nabla^2 \bar{\theta}; \phi_k^{\theta} \big\rangle, \\
        \mathfrak{L}^{(\omega,\psi)}_{i,k} &= \big\langle \dfrac{1}{\text{Re}} \nabla^2 \phi_i^{\omega} - J(\phi_i^{\omega},\bar{\psi}) - J(\bar{\omega},\phi_i^{\psi}) ; \phi_k^{\omega} \big\rangle, \\
        \mathfrak{L}^{(\omega,\theta)}_{i,k} &= \big\langle \text{Ri} \dfrac{\phi_i^{\theta}}{\partial x} ; \phi_k^{\omega} \big\rangle, \\
        \mathfrak{L}^{(\theta,\psi)}_{i,k} &= \big\langle -J(\bar{\theta},\phi_i^{\psi}) ; \phi_k^{\theta} \big\rangle, \\
        \mathfrak{L}^{(\psi,\theta)}_{i,k} &= \big\langle \dfrac{1}{\text{Re}\text{Pr}}\nabla^2 \phi_i^{\theta} -J(\phi^{\theta},\bar{\psi}) ; \phi_k^{\theta} \big\rangle, \\
        \mathfrak{N}^{(\omega,\psi)}_{i,j,k} & = \big\langle -J(\phi_i^{\omega},\phi_j^{\psi}) ; \phi_k^{\omega} \big\rangle, \\
        \mathfrak{N}^{(\theta,\psi)}_{i,j,k} & = \big\langle -J(\phi_i^{\theta},\phi_j^{\psi}) ; \phi_k^{\theta} \big\rangle. 
    \end{align*}
For all cases, the initial condition to initiate the ODE solver are obtained by projecting the initial field (mean-subtracted) onto the POD space of the first sub-interval as
\begin{align} 
\alpha^{(p=1)}_k(t_0) &= \big\langle u(\mathbf{x}, t_0) - \bar{u}^{(p=1)}(\mathbf{x})  ; \phi^{(p=1)}_k(\mathbf{x}) \big\rangle. \label{eq:init}
\end{align}
The only remaining part to close this section is to determine how to update the working manifold at the interface when moving from the $p$-th interval to the $(p+1)$-th interval (i.e., when $t = t_{\kappa^{(p)}}$). Once ROM solver reaches the end of the current interval, a reconstruction back to FOM space should be done. Subsequently, this FOM field is projected back on the new basis functions. These two steps can be summarized as follows,
\begin{align*}
    &(1) \qquad u(\mathbf{x}, t_{\kappa^{(p)}}) = \bar{u}^{(p)}(\mathbf{x}) + \sum_{k=1}^{R} \alpha_k^{(p)}(t_{\kappa^{(p)}}) \phi^{(p)}(\mathbf{x}), \\
    &(2) \qquad \alpha_k^{(p+1)}(t_{\kappa^{(p)}}) =  \big\langle u(\mathbf{x}, t_{\kappa^{(p)}})  - \bar{u}^{(p+1)}(\mathbf{x})  ; \phi_k^{(p+1)} \big\rangle,
\end{align*}
where the update step (manifold transfer) can be written as
\begin{align}
    \alpha_k^{(p+1)}(t_{\kappa^{(p)}}) &= \big\langle \bar{u}^{(p)}(\mathbf{x}) - \bar{u}^{(p+1)}(\mathbf{x}) ; \phi_k^{(p+1)} \big\rangle \nonumber \\ 
    & + \big\langle \sum_{k=1}^{R} \alpha_k^{(p)}(t_{\kappa^{(p)}}) \phi^{(p)}(\mathbf{x})   ; \phi_k^{(p+1)} \big\rangle. \label{eq:update}
\end{align}
This allows us to re-initiate our solver at the first timestep of the new time interval. Mathematically, this corresponds to imposing the following condition at the interface \cite{dihlmann2011model},
\begin{equation}
    \big\langle u^{(p)}(\mathbf{x}, t_{\kappa(p)}) - u^{(p+1)}(\mathbf{x}, t_{\kappa(p)}) ; \phi_k^{(p+1)} \big\rangle = 0.
\end{equation}

\section{Non-intrusive reduced order modeling}
\label{sec:LSTM}
\begin{figure*}[!ht]
	\centering
	\includegraphics[width=0.95\textwidth]{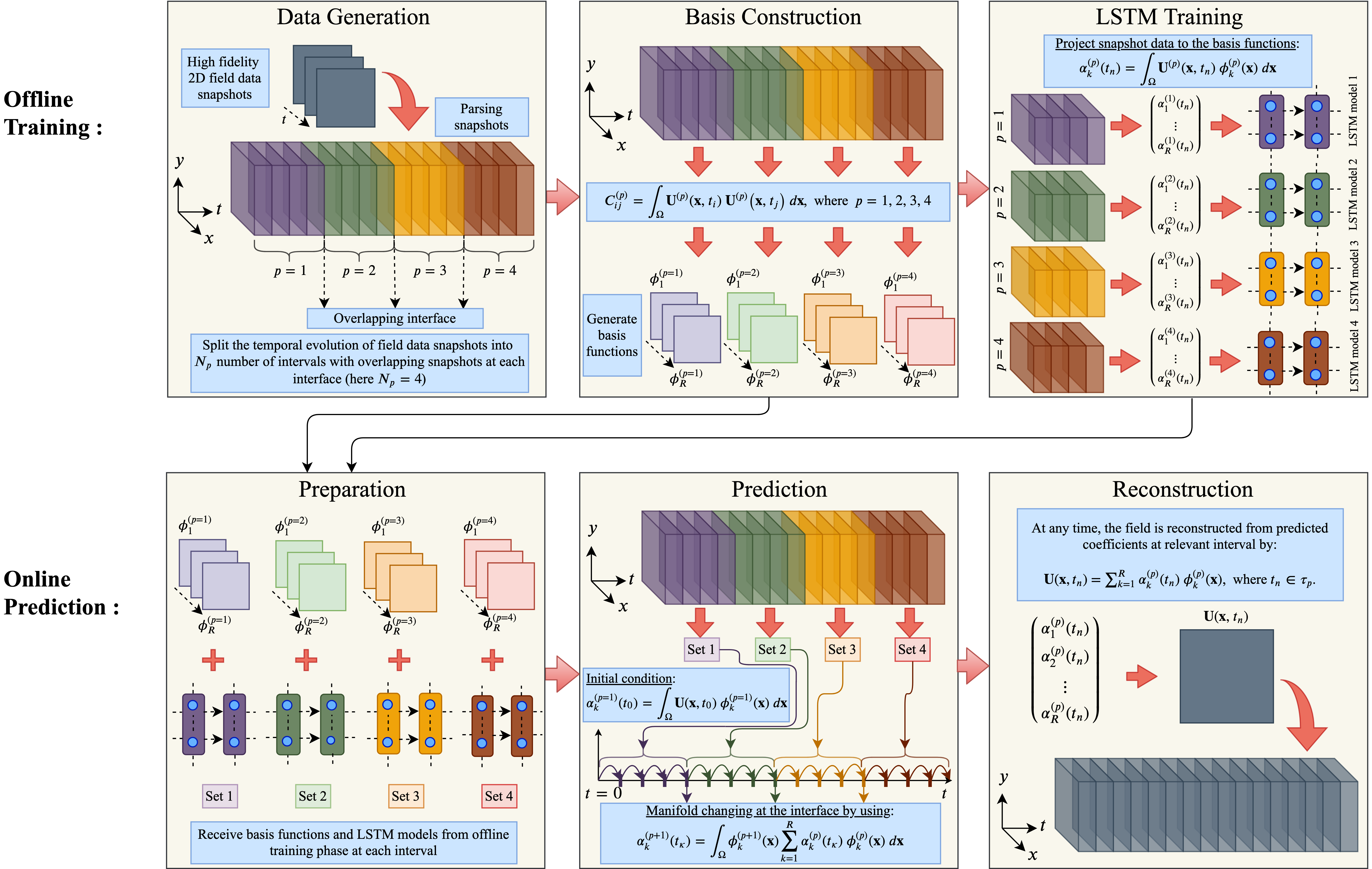}
	\caption{The proposed non-intrusive principal interval decomposition LSTM framework for unsteady non-ergodic flows.}
	\label{fig:pid}
\end{figure*}

In this section, we devise the proposed non-intrusive PID-LSTM framework for unsteady convective flows. To illustrate the PID-LSTM framework, we depict a workflow schematic diagram in Figure~\ref{fig:pid} for any arbitrary two-dimensional unsteady flow problem. In our two-dimensional representation of the PID-LSTM framework, $\mathbf{U}$ denotes any arbitrary two-dimensional field, for example, $\omega$ in Navier-Stokes and $\theta$ in Boussinesq test problems discussed in the present study. However, with proper modification in the LSTM architecture, this framework can be utilized for any three-dimensional field data and one-dimensional field data, for example, the 1D Burgers case in our study. Also, this PID-LSTM framework is parallelization friendly (e.g., using  parareal framework\cite{gander2007analysis,carlberg2019data}). As shown in Figure~\ref{fig:pid}, the first two stages of offline training phase in the PID-LSTM framwework is similar to the intrusive PID-GP framework described in Section~\ref{sec:GP} that we first split the stored high-fidelity field data snapshots into desired number of intervals and then generate the basis functions as well as the modal coefficients locally for each interval (i.e., using Equation~\ref{eq:upid2}). 

For ROM dynamics, we replace the Galerkin projection of PID-GP approach with LSTM recurrent neural network (RNN) architecture to make the framework fully non-intrusive or data-driven. There have been a number of research efforts which showed that RNN, specifically LSTM as a variant of RNN, is capable of predicting the dependencies among temporal data sequences \cite{yeo2019deep,jaeger2004harnessing,lecun2015deep,vlachas2018data,mohan2018deep,sak2014long}. Hence, we utilize the LSTM neural network architecture to model and predict the time-varying modal coefficient data sequences for our non-intrusive ROM framework. In our PID-LSTM formulation, we train individual LSTM architectures for the modal coefficients from each intervals which gives us the individual LSTM model for the respective PID interval. It should be noted that training the LSTM architectures is the computationally heavier part of the overall framework which is done in the offline phase (the top-right box in Figure~\ref{fig:pid}). Before describing the online testing phase, we briefly describe the LSTM architecture utilized in our study.

The standard RNN architecture suffers from issues like vanishing gradient problem \cite{bengio1994learning} which leads to the development of improved RNN architectures. LSTM is one of the most successful upgrades of traditional RNN architecture which can learn and predict the temporal dependencies between the given data sequence based on the input information and previously acquired information \cite{Gers99learningto,hochreiter1997long}. The conventional LSTM operates by the cell states stored in the memory blocks and gating mechanisms to control the flow of information. Each  memory  block  has  an input gate controlling the flow of input activations into the cell, a forget gate to adaptively forgetting and resetting the cell’s memory (to prevent over-fitting by processing continuous inflow of input streams), and the output gate controlling the output flow of cell activations into the next cell. To demonstrate our LSTM architecture for the present study, we can focus on the first interval only where the input sequential data matrix for training can be denoted as $\mathcal{X}_k$ and the output sequential data matrix $\mathcal{Y}_k$. Each sample of the input training matrix $\mathcal{X}_k$ is constructed as $\big \{\alpha_{1}^{(1)} (t_{n-\eta+1}), \dots , \alpha_{R}^{(1)} (t_{n-\eta+1}); \allowbreak  \dots ; \alpha_{1}^{(1)} \left(t_{n-1}\right), \dots , \alpha_{R}^{(1)} \left(t_{n-1}\right) ; \allowbreak \alpha_{1}^{(1)} \left(t_{n}\right), \dots , \alpha_{R}^{(1)} \left(t_{n}\right)\big \} $ and the corresponding output sample in output sequential data matrix $\mathcal{Y}_k$ is $\left\{\alpha_{1}^{(1)} (t_{n+1}) ,.., \alpha_{R}^{(1)} \left(t_{n+1}\right)\right\}$. Here, $\eta$ is the time history over which the LSTM model does the training and prediction recursively, called number of lookbacks. In our study, a constant value of $\eta$ is set 5 for the test cases to avoid complexity while analyzing the results. 

\begin{figure}[!ht]
	\centering
	\includegraphics[width=0.5\textwidth]{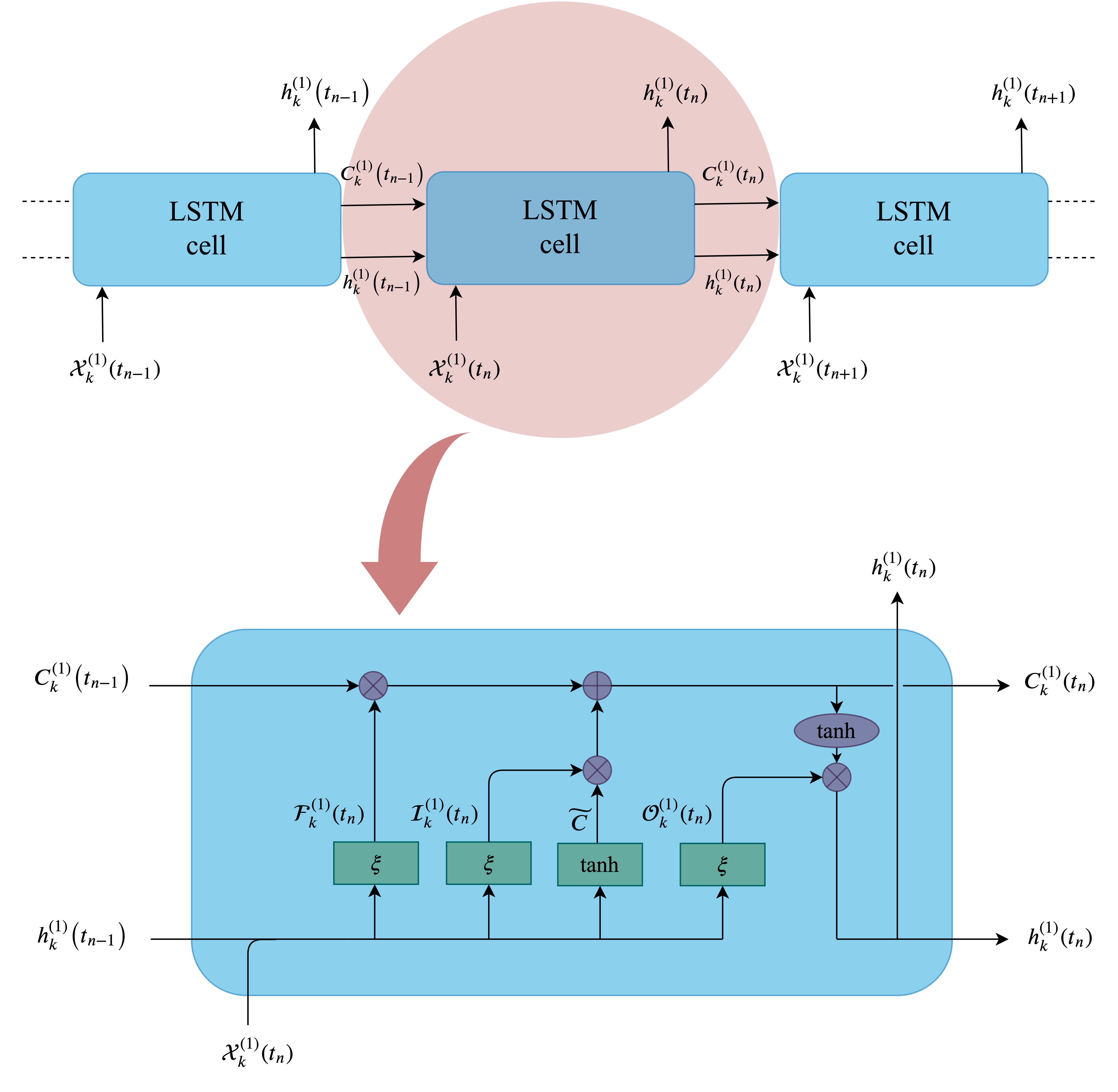}
	\caption{Schematic of a typical LSTM cell unit.}
	\label{fig:lstmcell}
\end{figure}

To illustrate the data stream flow through a standard LSTM cell, we have shown the sketch of a typical LSTM cell unit in Figure~\ref{fig:lstmcell}. The basic LSTM equations to compute the gate functions can be given by:
\begin{align}
    &\mathcal{F}_k^{(1)} (t_n) =  \xi \left(W_{fh} h_k^{(1)} (t_{n-1}) + W_{f\mathcal{X}} \mathcal{X}_k^{(1)} (t_n) + b_f\right),  \label{pidlstm1} \\
    &\mathcal{I}_k^{(1)} (t_n) =  \xi \left(W_{ih} h_k^{(1)} (t_{n-1}) + W_{i\mathcal{X}} \mathcal{X}_k^{(1)} (t_n) + b_i\right), \label{pidlstm2}\\
    &\mathcal{O}_k^{(1)} (t_n) =  \xi \left(W_{oh} h_k^{(1)} (t_{n-1}) + W_{o\mathcal{X}} \mathcal{X}_k^{(1)} (t_n) + b_o\right), \label{pidlstm3}
\end{align}
where $\mathcal{I}$, $\mathcal{F}$ and $\mathcal{O}$ represents the input, forget and output gates, respectively. The LSTM cell output activation vector or the hidden state vector is denoted as $h$, while $\xi$ represents the logistic sigmoid function, $b$ denotes the bias vectors and $W$ represents the weight matrices for each gates. Assuming the cell activation vector or internal cell state vector as $C$, the internal cell state equation can be expressed as:
\begin{flalign}\label{pidlstm4}
    C_k^{(1)} (t_n) =  \mathcal{F}_k^{(1)} (t_n)\odot C_k^{(1)} (t_{n-1}) + \mathcal{I}_k^{(1)} (t_n) \odot \widetilde{C},
\end{flalign}
where $\widetilde{C} =  \text{tanh}\left(W_{ch} h_k^{(1)} (t_{n-1}) + W_{c\mathcal{X}} \mathcal{X}_k^{(1)} (t_n) + b_c\right)$ and $\odot$ is the Hadamard product of two vectors. The output state of each LSTM cell is given by:
\begin{flalign}\label{pidlstm5}
    h_k^{(1)} (t_n) =  \mathcal{O}_k^{(1)} (t_n) \odot \text{tanh}\left(C_k^{(1)} (t_n)\right).
\end{flalign}
We utilize Keras API to design the LSTM architecture for our PID-LSTM framework \cite{chollet2015keras}. 
The hyperparameters that we used in our numerical experiments implementing PID-LSTM are listed in Table~\ref{tab:LSTMparameters}. The mean-squared error (MSE) is chosen as the loss function for weight-optimization, and a variant of stochastic gradient descent method, called ADAM \cite{kingma2014adam}, is used to optimize the mean-squared loss. We utilize tanh activation function in each LSTM layer. The batch size and epochs are set to 64 and 100 respectively. We select randomly $20 \%$ of the training data for validation during training. We maintain a constant hyperparameter set up to fairly compare the results for different numerical experiments. The training data is normalized by the minimum and maximum of each time series to be in between the range $[-1, +1]$.

In the online testing phase, we use the generated basis functions and LSTM models for each intervals from the given snapshot data to recursively predict the coefficients until final time. For testing, the input of the first trained model will be the initial states of the first interval (see Equation~\ref{eq:init}).
When the online prediction reaches the end of the sub-interval, a manifold transfer is done at the interface by following Equation~\ref{eq:update} and the framework switches to the next LSTM model. 
By performing this procedure recursively, the modal coefficients until final time can be determined and then the field can be reconstructed at any time from the predicted coefficients at relevant interval by using Equation~\ref{eq:upid}. Because the LSTM framework can have a wider interface than PID-GP framework (i.e., for $\eta>1$), a buffer zone can be defined at the interface, where Equation~\ref{eq:update} is applied for $t_{\kappa^{(p)}}, t_{\kappa^{(p)}-1}, \dots, t_{\kappa^{(p)}-\eta+1} $. This would be the input for the subsequent LSTM model to predict $\alpha\left(t_{\kappa^{(p)}+1}\right)$.

\begin{table*}[htbp]
    \centering
    \caption{A list of hyperparameters utilized to train the LSTM network for all numerical experiments.}
    \begin{tabular}{p{0.38\textwidth}p{0.15\textwidth}p{0.15\textwidth}p{0.15\textwidth}p{0.12\textwidth}}
    \hline\noalign{\smallskip}
    Variables & \quad \quad \ Burgers & Vortex merger & Double shear layer & \quad Boussinesq\\
    \noalign{\smallskip}\hline\noalign{\smallskip}
    Number of hidden layers  & \quad \quad \ \ 3 & \ 3 & \ 4 & \quad \ 3\\
    Number of neurons in each hidden layer  & \quad \quad \ \ 80 & \ 80 & \ 80 & \quad \ 80\\
    Batch size & \quad \quad \ \ 64 & \ 64 & \ 64 & \quad \ 64\\
    Epochs & \quad \quad \ \ 100 & \ 1000 & \ 1000 &  \quad \ 100\\
    Activation functions in the LSTM layers & \quad \quad \ \ tanh & \ tanh & \ tanh & \quad \ tanh\\
    Validation data set & \quad \quad \ \ 20$\%$ & \  20$\%$ & \ 20$\%$ & \quad \ 20$\%$\\
    Loss function &  \quad \quad \ \ MSE & \ MSE & \ MSE & \quad \ MSE\\
    Optimizer & \quad \quad \ \ ADAM & \ ADAM & \ ADAM & \quad \ ADAM\\
    Learning rate & \quad \quad \ \ 0.001 & \ 0.001 & \ 0.001 & \quad \ 0.001\\
    First moment decay rate & \quad \quad \ \ 0.9 & \ 0.9 & \ 0.9 & \quad \ 0.9\\
    Second moment decay rate & \quad \quad \ \ 0.999 & \ 0.999 & \ 0.999 & \quad \ 0.999\\
    \noalign{\smallskip}\hline
    \end{tabular}
    \label{tab:LSTMparameters}
\end{table*}

\section{Results}
\label{sec:results}
For all test cases, we applied both the intrusive and non-intrusive frameworks, sketched in Sections~\ref{sec:GP} and \ref{sec:LSTM}, respectively. A total number of $800$ snapshots, or strobes, were collected from FOM simulations. A summary of the data generation characteristics as well as CPU time for constructing basis functions, implemented in FORTRAN, is given in Table~\ref{tab:FOMparameters}. For numerical computations, we used a family of fourth order compact schemes for spatial derivatives \cite{lele1992compact} and a third order Runge-Kutta scheme for temporal integration \cite{gottlieb1998total}. The time domain is decomposed into 2, 4, 8, and 16 subintervals. The CPU times of PID-LSTM training and testing stages (in Python) are summarized in Table~\ref{tab:cpu} for different number of intervals. It should be noted that the most expensive stages of PID-LSTM approach is performed offline, where the online predictions is relatively fast. Although the PID increases the computational overhead for online prediction (almost linearly with number of intervals), speedups of several orders of magnitudes compared to FOM are accomplished. Prediction and reconstruction accuracies are computed and compared with the case without interval decomposition (i.e., a single global interval) in order to illustrate the effects of non-ergodicity and assess the PID contribution to mitigate those effects. Also, the predictive performance of LSTM-based approach is compared with the GP-based one to demonstrate its capability to bypass the Galerkin-projection step and provide accurate predictions without prior information of the complex physical information.

\begin{table*}[htbp]
    \centering
    \caption{A summary of data generation characteristics for the full order models and their required CPU times (in seconds). We document the corresponding speedup reached by the PID-GP model. We note that the CPU time assessments documented in this table are based on FORTRAN executions.}
    \begin{tabular}{p{0.4\textwidth}p{0.14\textwidth}p{0.14\textwidth}p{0.14\textwidth}p{0.14\textwidth}}
    \hline\noalign{\smallskip}
    Variables & \quad \quad \ Burgers &  Vortex merger & Double shear layer & \quad Boussinesq\\
    \noalign{\smallskip}\hline\noalign{\smallskip}
    \multicolumn{5}{l}{\emph{\underline{FOM relevant parameters}}} \smallskip \\
    Grid resolution & \quad \quad \ \ 1024 & \ $1024 \times 1024$ & \ $1024 \times 1024$ & \quad \ $4096 \times 512$\\
    Time step, $\Delta t$ & \quad \quad \ \ $1.00\text{E}-4$ & \ $1.00\text{E}-3$ & \ $1.00\text{E}-3$ & \quad \ $5.00\text{E}-4$\\
    Maximum simulation time, $t_{\text{max}}$ & \quad \quad \ \ 2.00 & \ 40.00 & \ 40.00 & \quad \ 8.00\\
    CPU time required for FOM simulation  & \quad \quad \ \ $3.43$ & \ $1.20\text{E}5$ & \ $1.20\text{E}5$  & \quad \ $7.38\text{E}4$  \smallskip \\      
    \multicolumn{5}{l}{\emph{\underline{Offline data preparation}}} \smallskip \\
    Number of snapshots collected, $N_s$ & \quad \quad \ \ 800 & \  800 & \ 800 & \quad \ 800\\
    CPU time for generating POD bases ($N_p = 1$) & \quad \quad \ \ $7.21\text{E}1$ & \ $3.31 \text{E}4$ & \ $3.27\text{E}4$ & \quad \ $1.18\text{E}5$  \\
    CPU time for generating POD bases ($N_p = 2$) & \quad \quad \ \ $7.70$ & \ $8.15\text{E}3$ & \ $8.17\text{E}3$  & \quad \ $2.93\text{E}4$ \\
    CPU time for generating POD bases ($N_p = 4$) & \quad \quad \ \ $9.33\text{E}-1$ & \ $2.15\text{E}3$ & \ $2.15\text{E}3$ & \quad \ $7.64\text{E}3$\\
    CPU time for generating POD bases ($N_p = 8$) & \quad \quad \ \ $1.53\text{E}-1$ & \ $5.70\text{E}2$ & \ $5.70\text{E}2$ & \quad \ $1.96\text{E}3$ \\
    CPU time for generating POD bases ($N_p = 16$) & \quad \quad \ \ $2.98\text{E}-2$ & \ $1.22\text{E}2$  & \ $1.25\text{E}2$ & \quad \ $4.55\text{E}2$ \smallskip \\
    \multicolumn{5}{l}{\emph{\underline{PID-GP online phase}}} \smallskip \\
    Number of modes retained, $R$ & \quad \quad \ \ 6 & \ 6 & \ 6 & \quad \ 6\\
    CPU time for PID-GP [speedup] ($N_p = 1$) & \quad \quad \ \ $0.14$  [25] & \ $0.03$ [$4\times10^6$] & \ $0.03$ [$4\times10^6$] & \quad \ $0.06$ [$1\times10^6$]  \\
    CPU time for PID-GP [speedup] ($N_p = 2$) & \quad \quad \ \ $0.13$ [26] & \ $0.13$ [$9\times10^5$] & \ $0.12$ [$1\times10^6$]  & \quad \ $0.43$ [$2\times10^5$] \\
    CPU time for PID-GP [speedup] ($N_p = 4$) & \quad \quad \ \ $0.11$ [31] & \ $0.30$ [$4\times10^5$] & \ $0.32$ [$4\times10^5$] & \quad \ $1.16$ [$6\times10^4$] \\
    CPU time for PID-GP [speedup] ($N_p = 8$) & \quad \quad \ \ $0.07$ [49] & \ $0.63$ [$2\times10^5$] & \ $0.64$ [$2\times10^5$] & \quad \ $2.61$ [$3\times10^4$] \\
    CPU time for PID-GP [speedup] ($N_p = 16$) & \quad \quad \ \ $0.11$ [31] & \ $1.27$ [$9\times10^4$]  & \ $1.27$ [$9\times10^4$] & \quad \ $4.97$ [$1\times10^4$] \\
    \noalign{\smallskip}\hline
    \end{tabular} 
    \label{tab:FOMparameters}
\end{table*}

\begin{table}[htbp]
	\centering
	\caption{CPU time (in seconds) of PID-LSTM training (offline) and testing (online) stages. We note that the CPU time assessments documented in this table are based on Python executions.}
	\begin{tabular}{p{0.15\textwidth}p{0.15\textwidth}p{0.15\textwidth}} 
		\hline\noalign{\smallskip} 
		$N_p$ & Training time &   Testing time    \\ 
		\noalign{\smallskip}\hline\noalign{\smallskip}
		\multicolumn{3}{l}{\emph{\underline{1D Burgers}}}\smallskip \\
		$1$ & \ $17.62$ & \ $2.04$  \\
        $2$ & \ $30.18$ & \ $1.98$  \\
        $4$ & \ $61.88$ & \ $3.78$  \\
        $8$ & \ $132.70$ & \ $11.28$  \\
        $16$& \ $304.40$ & \ $41.09$   \medskip \\ 
		\multicolumn{3}{l}{\emph{\underline{2D Vortex Meger}}}\smallskip \\
		$1$ & \ $125.51$ & \ $2.03$  \\
        $2$ & \ $248.57$ & \ $4.52$  \\
        $4$ & \ $533.28$ & \ $11.00$  \\
        $8$ & \ $1399.11$ & \ $35.80$  \\
        $16$& \ $2379.21$ & \ $79.74$   \medskip \\ 
		\multicolumn{3}{l}{\emph{\underline{2D Double Shear Layer}}}\smallskip \\
        $1$ & \ $161.76$ & \ $2.40$  \\
        $2$ & \ $333.11$ & \ $5.51$  \\
        $4$ & \ $809.51$ & \ $15.17$  \\
        $8$ & \ $1435.85$ & \ $31.36$  \\
        $16$& \ $3655.18$ & \ $94.85$   \medskip \\ 
		\multicolumn{3}{l}{\emph{\underline{2D Boussinesq}}}\smallskip \\
        $1$ & \ $17.36$ & \ $2.61$  \\
        $2$ & \ $36.01$ & \ $7.55$  \\
        $4$ & \ $82.86$ & \ $20.55$  \\
        $8$ & \ $238.22$ & \ $57.90$  \\
        $16$& \ $347.09$ & \ $116.52$   \\ 
		\hline
	\end{tabular}
	\label{tab:cpu}
\end{table}

\subsection{1D Burgers problem} 
\label{sec:BurgRes}
\begin{figure}[!ht]
	\centering
	\includegraphics[width=0.45\textwidth]{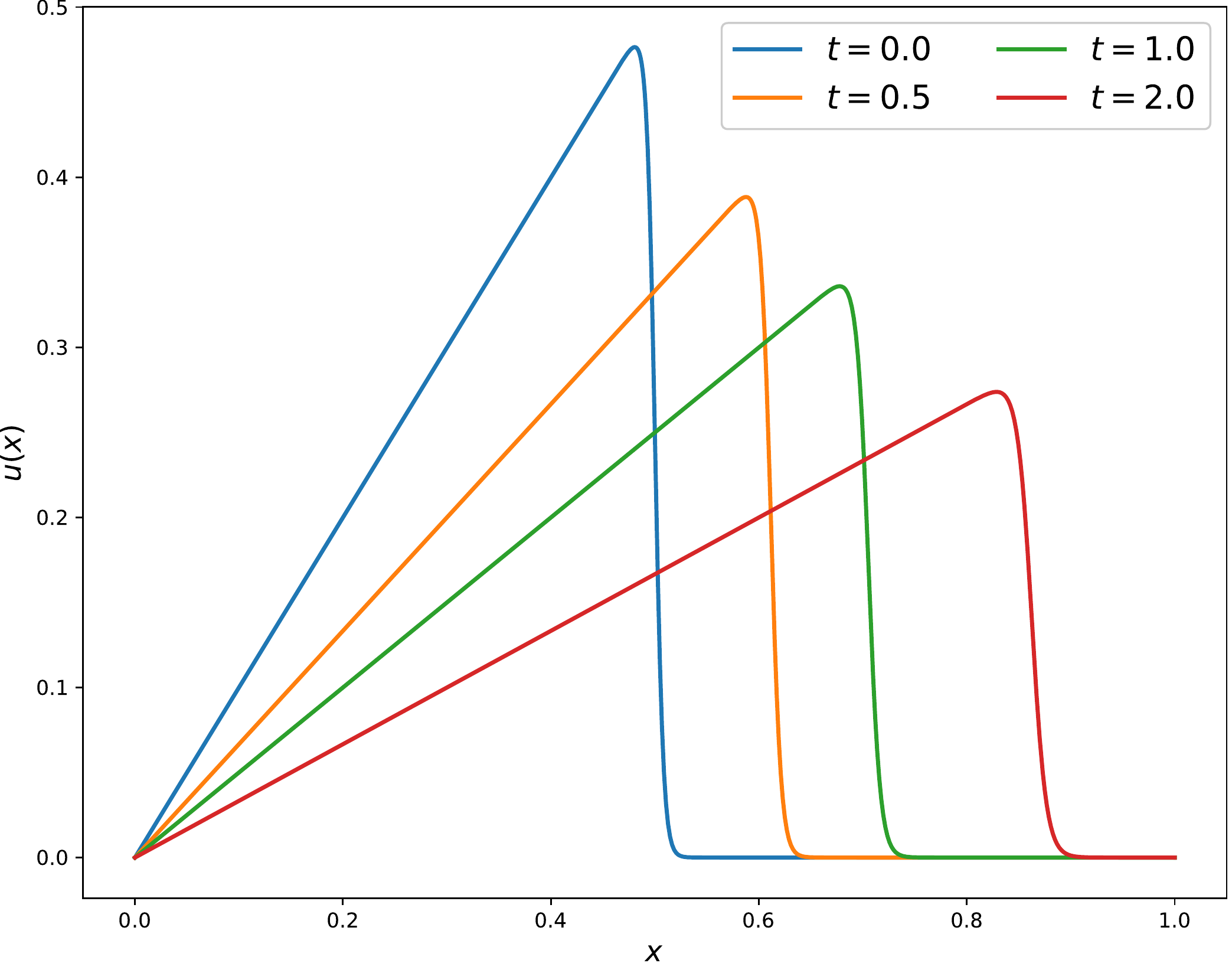}
	\caption{Velocity field at different time instances for Burgers equation for $\text{Re} = 1000$ using $1024$ grid and $\Delta t= 0.0001$.}
	\label{fig:Burgers}
\end{figure}

\begin{figure}[!hb]
	\centering
	\includegraphics[width=0.45\textwidth]{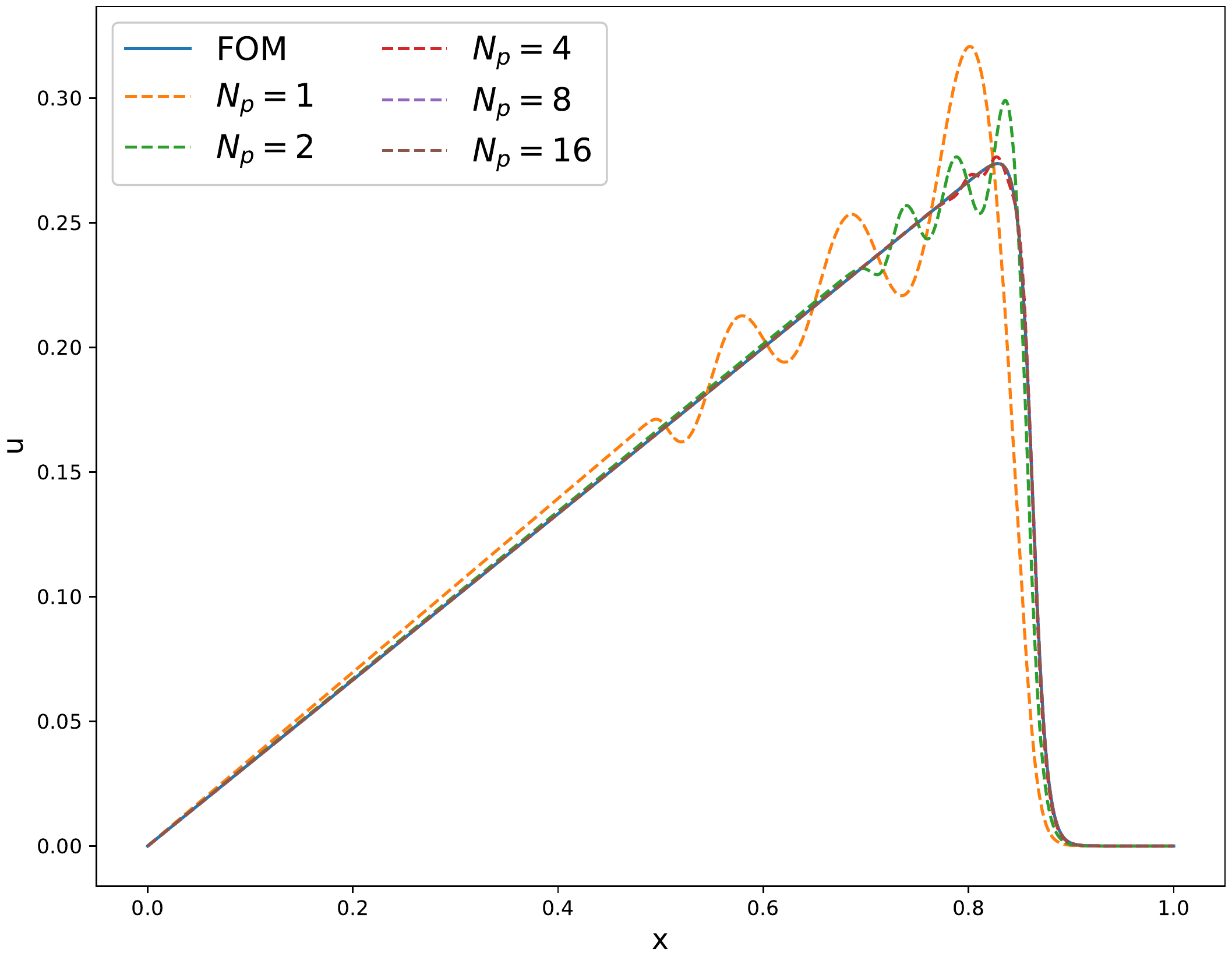}
	\caption{Final velocity field (i.e., at $t=2$) for Burgers problem from true projection using different number of intervals compared to FOM solution.}
	\label{fig:brg_POD}
\end{figure}

\begin{figure}[!ht]
	\centering
	\includegraphics[width=0.45\textwidth]{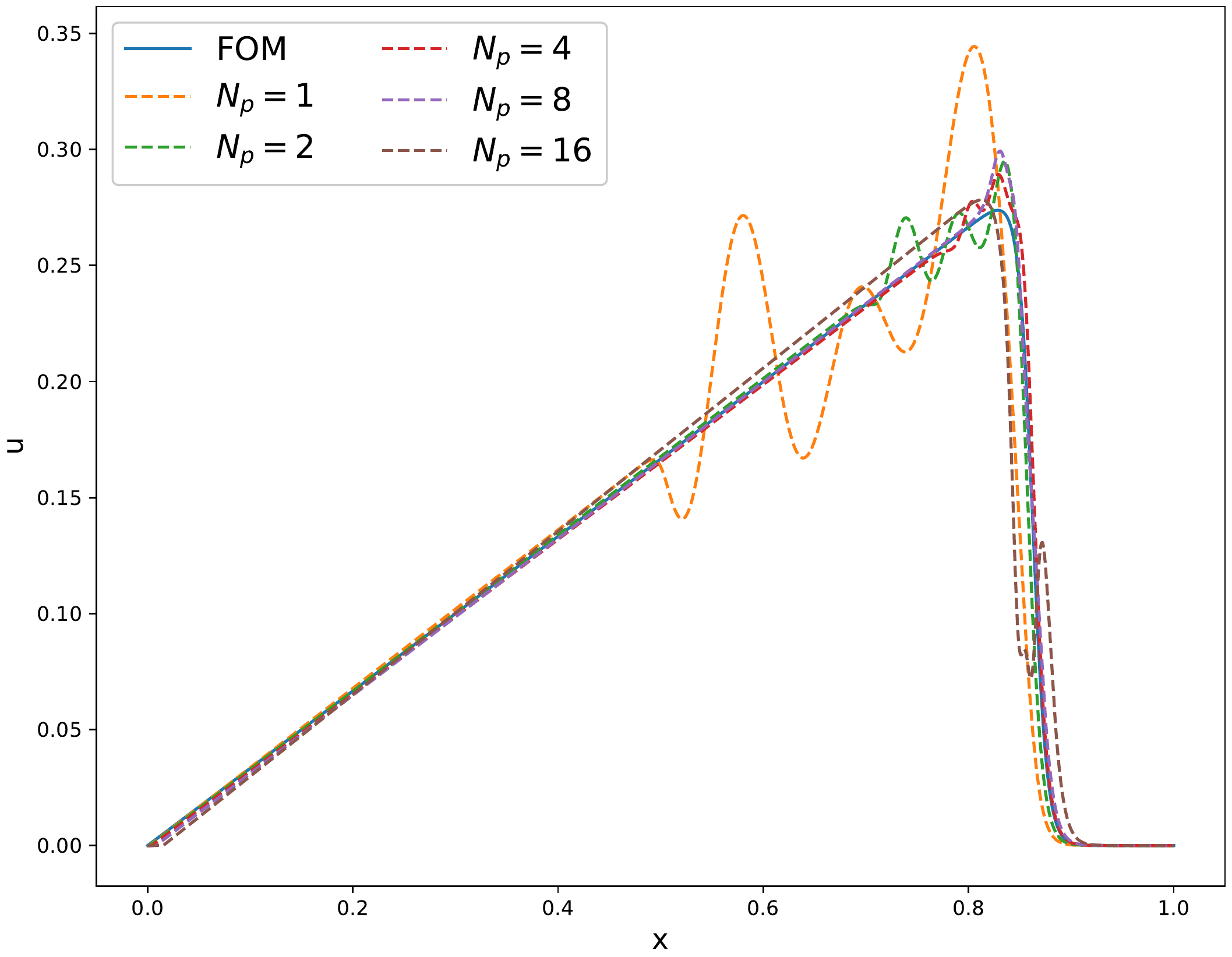}
	\caption{Final velocity field (i.e., at $t=2$) for Burgers problem from PID-GP prediction using different number of intervals compared to FOM solution.}
	\label{fig:brg_GP}
\end{figure}

\begin{figure}[!ht]
	\centering
	\includegraphics[width=0.45\textwidth]{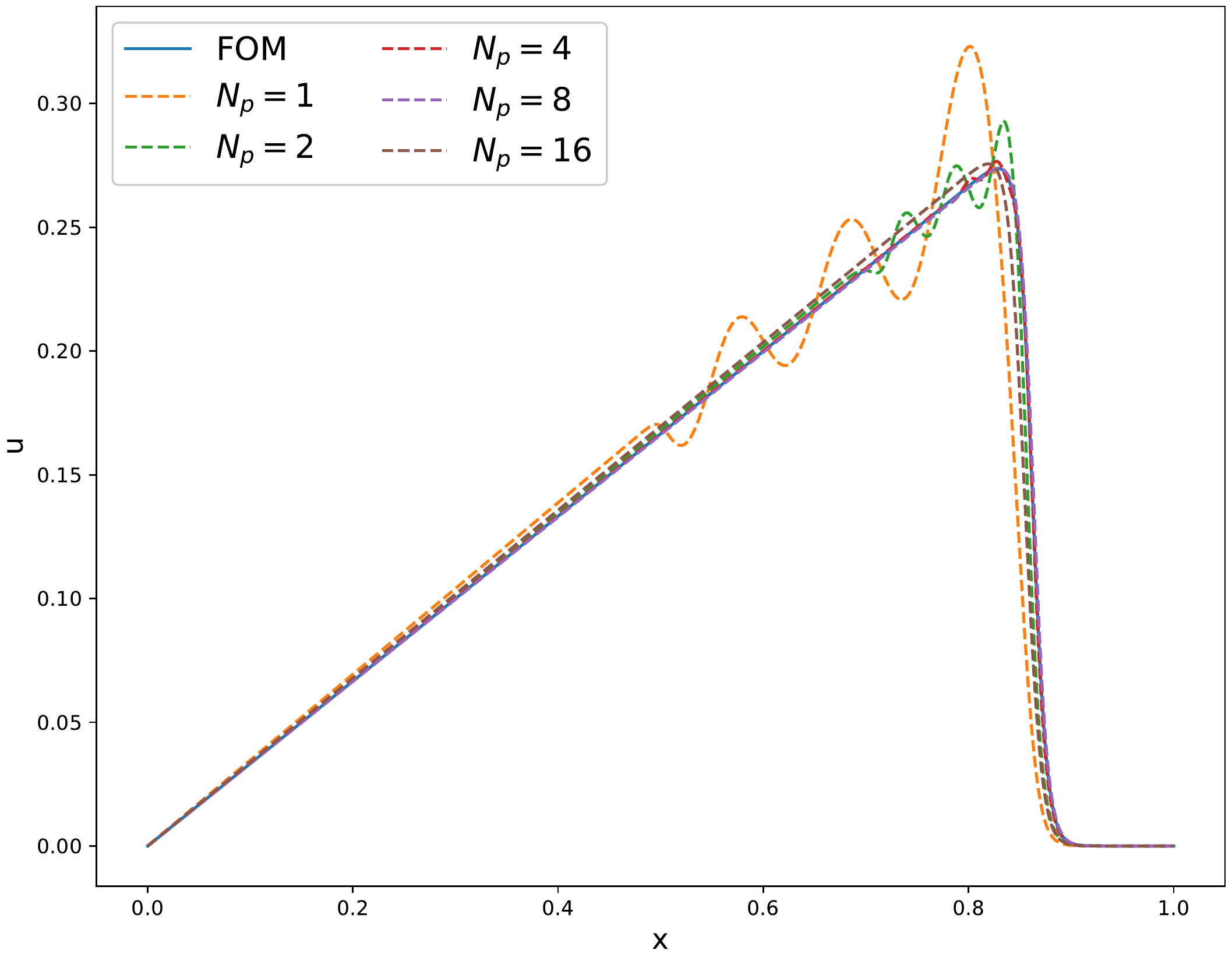}
	\caption{Final velocity field (i.e., at $t=2$) for Burgers problem from PID-LSTM prediction using different number of intervals compared to FOM solution.}
	\label{fig:brg_LSTM}
\end{figure}

\begin{figure}[!ht]
	\centering
	\includegraphics[width=0.45\textwidth]{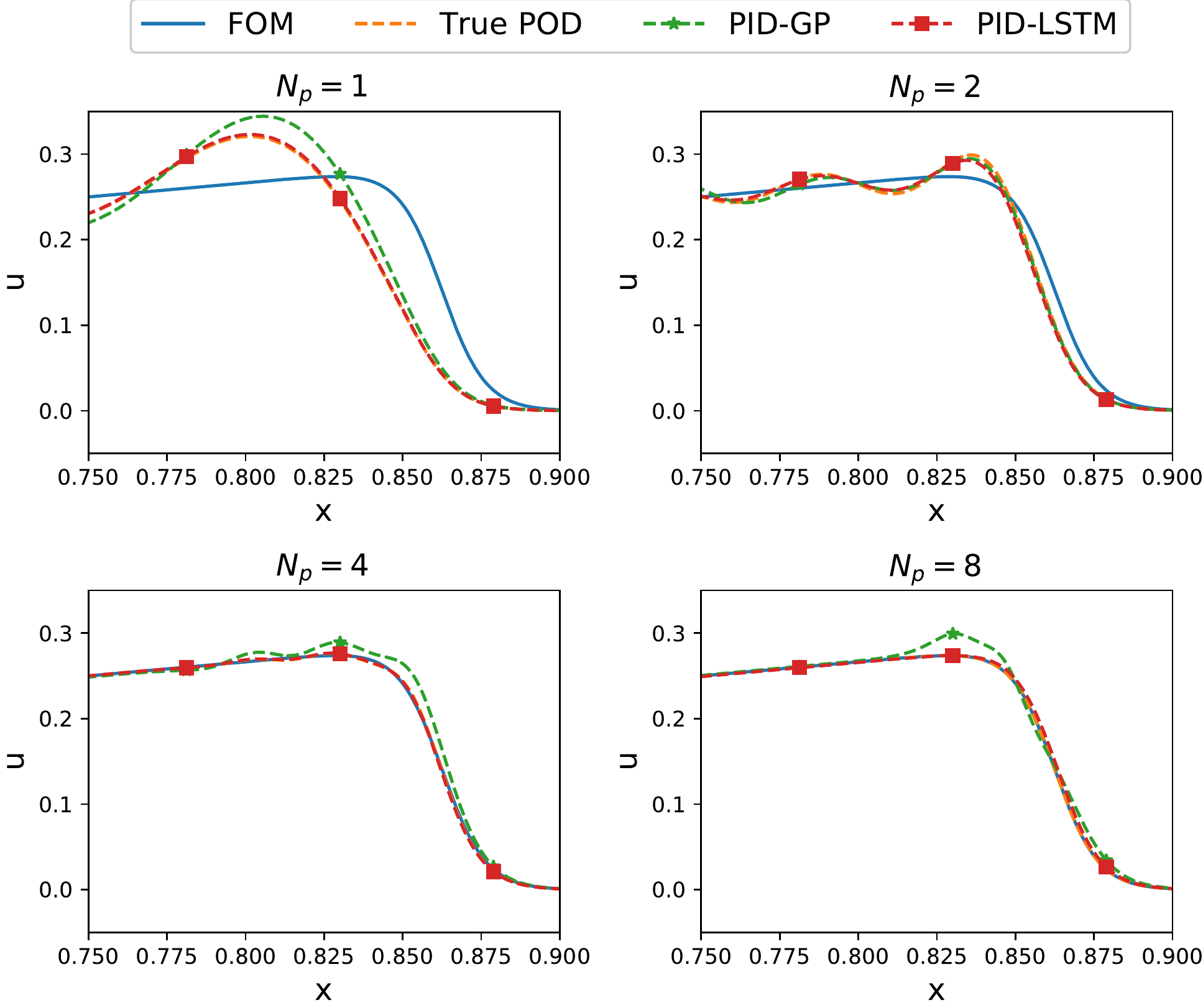}
	\caption{Comparison between final velocity field for Burgers problem obtained from different approaches using different number of intervals.}
	\label{fig:brg_comp}
\end{figure}

For 1D Burgers simulation, we consider the following initial condition \cite{maleewong2011line},
\begin{equation}
    u(x, 0) = \dfrac{x}{1 + \exp{\left( \dfrac{\text{Re}}{4} \left(x^2 - 1 \right)  \right)} },
\end{equation}
with $ x\in [0,1]$. Also, we assume Dirichlet boundary conditions where $u(0, t) = u(1,t) = 0$. We solved the problem for Reynolds number, $\text{Re} = 1000$ using $1024$ grid spacings in $x$-direction and a timestep of $10^{-4}$ for $t \in [0,2]$. We would like to mention here that the 1D Burgers equation with the above initial and boundary conditions has an analytic solution \cite{maleewong2011line}, but we preferred to solve it numerically for consistent comparison with ROMs. The solution $u(x, t)$ represents a traveling wave along a flat horizontal bottom, shown at different times in Figure~\ref{fig:Burgers}
In order to demonstrate the benefits of the PID approach constructing localized basis functions, we provide the true projection of the final field (at $t=2$), compared to the FOM solution in Figure~\ref{fig:brg_POD}. We can easily observe that using only a single global interval (i.e., $N_p=1$) gives inaccurate solution with oscillations that do not exist at any instance of the flow. This is due to the deformation and smoothing-out of the global modes by the rapidly evolving flow. The procedures of intrusive and non-intrusive ROMs ere applied to evaluate the temporal coefficients $\alpha(t)$. Results are shown in Figures~\ref{fig:brg_GP}-\ref{fig:brg_LSTM} with a zoomed-in view in Figure~\ref{fig:brg_comp}. Similar observations can be obtained regarding the effects of the number of intervals on the accuracy of the ROM approximation. However, it can be easily seen that the oscillations are amplified in PID-GP results compared to PID-LSTM. This is mainly due to the nonlinear interactions in the governing equation~\ref{eq:brg}, where the lower modes interacts strongly with the higher modes. Truncation of the lower modes simply ignores these interaction in projection-based ROM, Equation~\ref{eq:rombrg} and results in a deviation of the ROM dynamics from the true one.

\begin{figure}[!ht]
	\centering
	\includegraphics[width=0.45\textwidth]{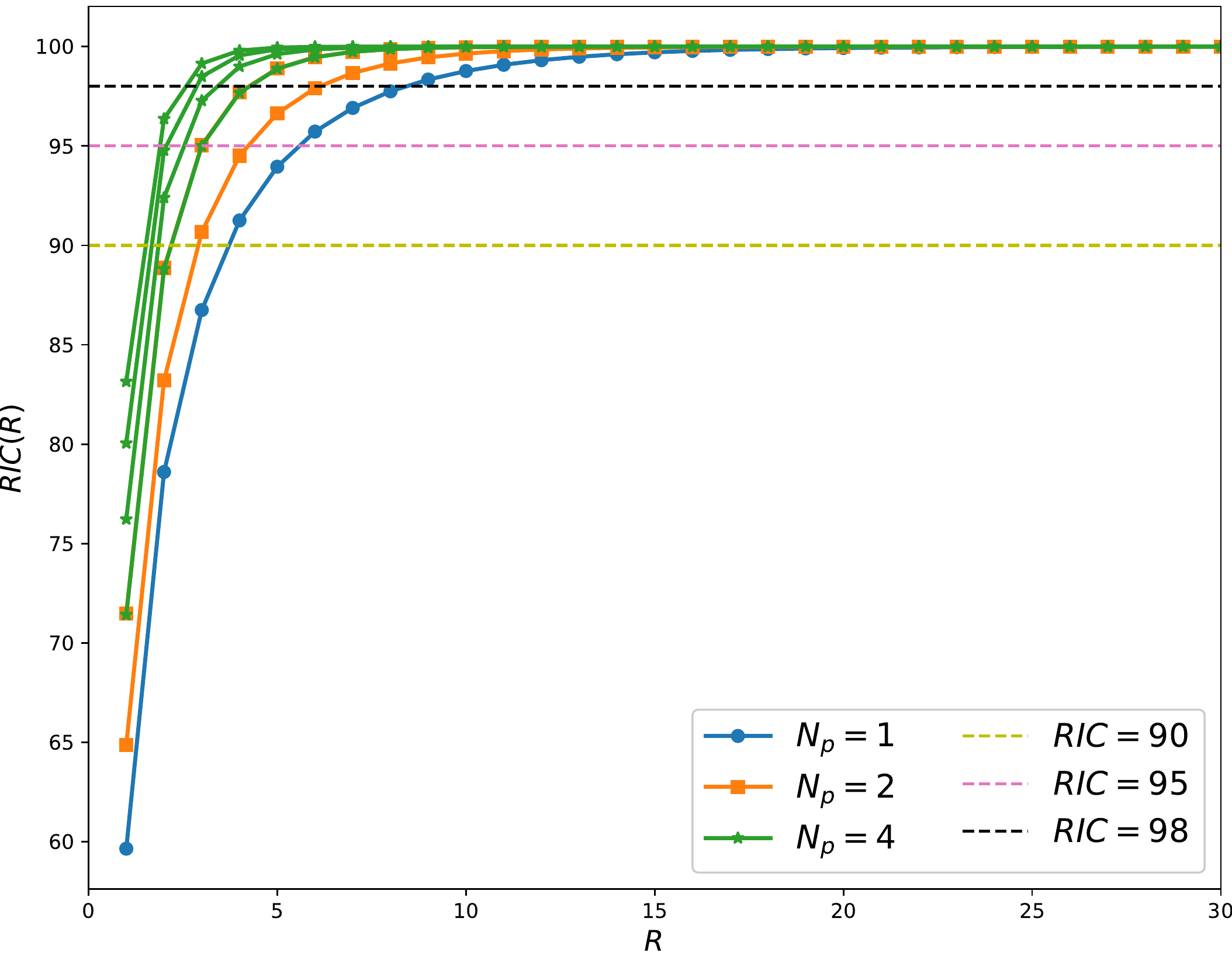}
	\caption{Relative information content ($RIC$) for Burgers problem, using one, two and four intervals.}
	\label{fig:brg_eig}
\end{figure}

A simple eigenvalue analysis of Burgers problem can help in demonstrating the idea behind interval decomposition. In POD, the percentage modal energy is computed using the following relative information content ($RIC$) formula \cite{gunzburger2012flow},
\begin{align}
	RIC(R) = \left(\frac{\sum_{j=1}^{R}\lambda_j} {\sum_{j=1}^{N_s}\lambda_j}\right) \times 100.
\end{align}
The $RIC$ plot is shown in Figure~\ref{fig:brg_eig} for different number of intervals. One can easily observe that this interval decomposition produces local POD modes with more concentrated energy content, compared to a single interval giving global modes with more distributed energies. For example, if we are interested in capturing $98 \%$ of the total energy (snapshots variance), we would need at least 8 POD modes in case of using a single wider interval. On the other hand, if we decompose our interval into two partitions, we will need 6 modes and if we decompose into 4 sub-intervals, 4 modes will be more than enough. Although this might imply more memory requirements, significant computational gains can be obtained. For instance, if we follow the classical POD-GP approach, the computational cost is $O(R^3)$. Therefore, using 4 modes instead of 8 modes would be around 8 times less costly.
\subsection{2D vortex merger problem} 
\label{sec:VortRes}
\begin{figure}[!ht]
	\centering
	\includegraphics[width=0.45\textwidth]{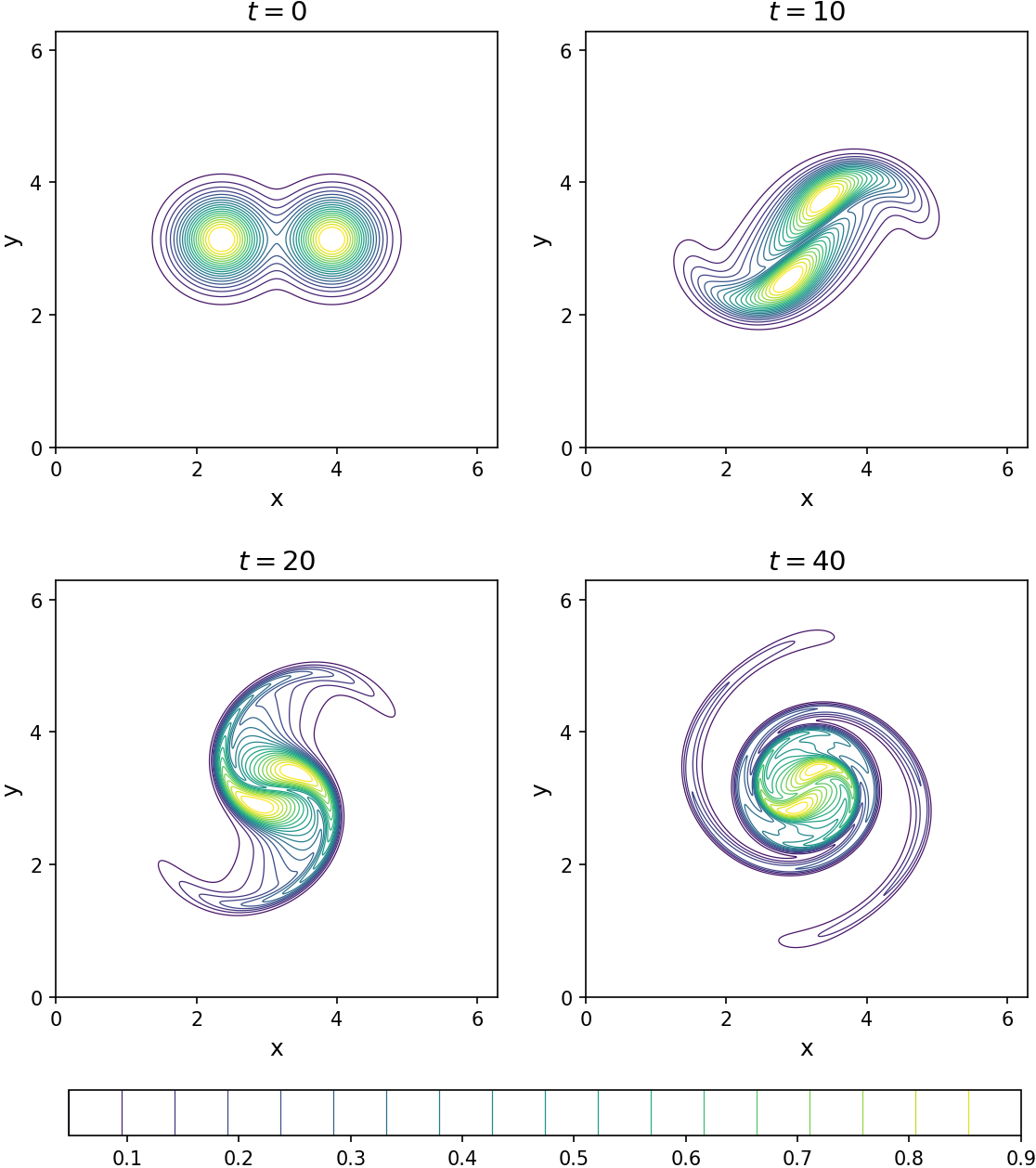}
	\caption{Vorticity field at different time instances for vortex merger problem using $1024^2$ grid and $\Delta t= 0.001$.}
	\label{fig:Vmerge}
\end{figure}

We expand our framework testing into two-dimensional cases. As an application for 2D Navier-Stokes equations is the vortex merger problem (i.e., the merging of co-rotating vortex pair) \cite{buntine1989merger}. The merging process occurs when two vortices of the same sign with parallel axes are within a certain critical distance from each other, ending as a single, nearly axisymmetric, final vortex \cite{von2000vortex}. It is a two-dimensional process and is one of the fundamental processes of fluid motion and occurs in many fields such as astrophysics, meteorology, and geophysics. For example, in two-dimensional turbulence, like-sign vortex merger is the main factor affecting the evolution of the vortex population \cite{von2000vortex}. Vortex merging also plays an important role in the context of aircraft trailing wakes \cite{meunier2005physics}. 
We consider an initial vorticity field of two Gaussian-distributed vortices with a unit circulation as follows,
\begin{align}
    \omega(x, y, 0) &= \exp\left( -\rho \left[ (x-x_1)^2  + (y-y_1)^2 \right] \right) \nonumber  \\
                    &+ \exp{\left( -\rho \left[ (x-x_2)^2 + (y-y_2)^2 \right] \right)},
\end{align}
where the vortices centers are initially located at $(x_1,y_1) = (\dfrac{3\pi}{4},\pi)$ and $(x_2,y_2) = (\dfrac{5\pi}{4},\pi)$. We use a Cartesian domain $(x,y) \in [0,2\pi] \times [0,2\pi]$, with a periodic boundary conditions. We did our simulations solving Equation~\ref{eq:NS} with Reynolds number of $10,000$ using $1024^2$ spatial grid and a timestep of $0.001$. The evolution of the two vortices from time $t=0$ to $t=40$ is shown in Figure~\ref{fig:Vmerge}. Details of the numerical schemes and computations can be found in a previous study \cite{san2013coarse}. 

\begin{figure}[!ht]
	\centering
	\includegraphics[width=0.45\textwidth]{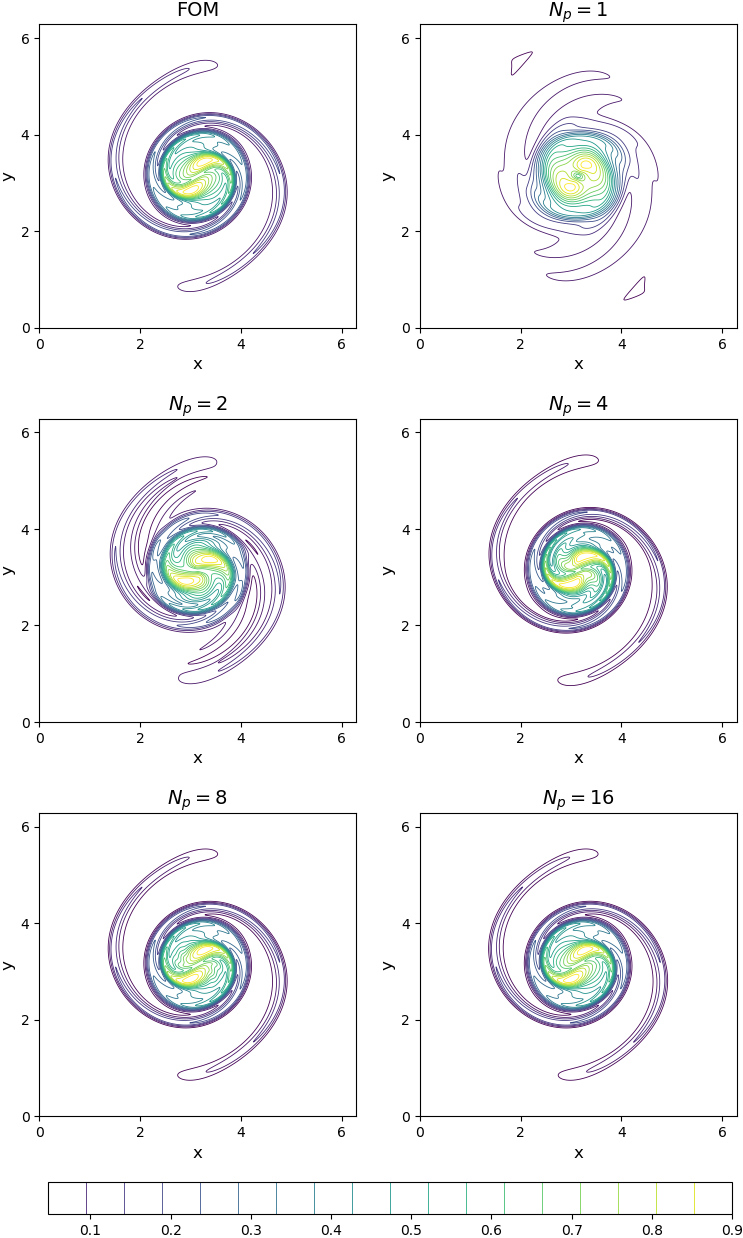}
	\caption{Final vorticity contours (i.e., at $t=40$) for vortex merger problem from true projection using different number of intervals compared to FOM solution.}
	\label{fig:VM_POD}
\end{figure}

\begin{figure}[!ht]
	\centering
	\includegraphics[width=0.45\textwidth]{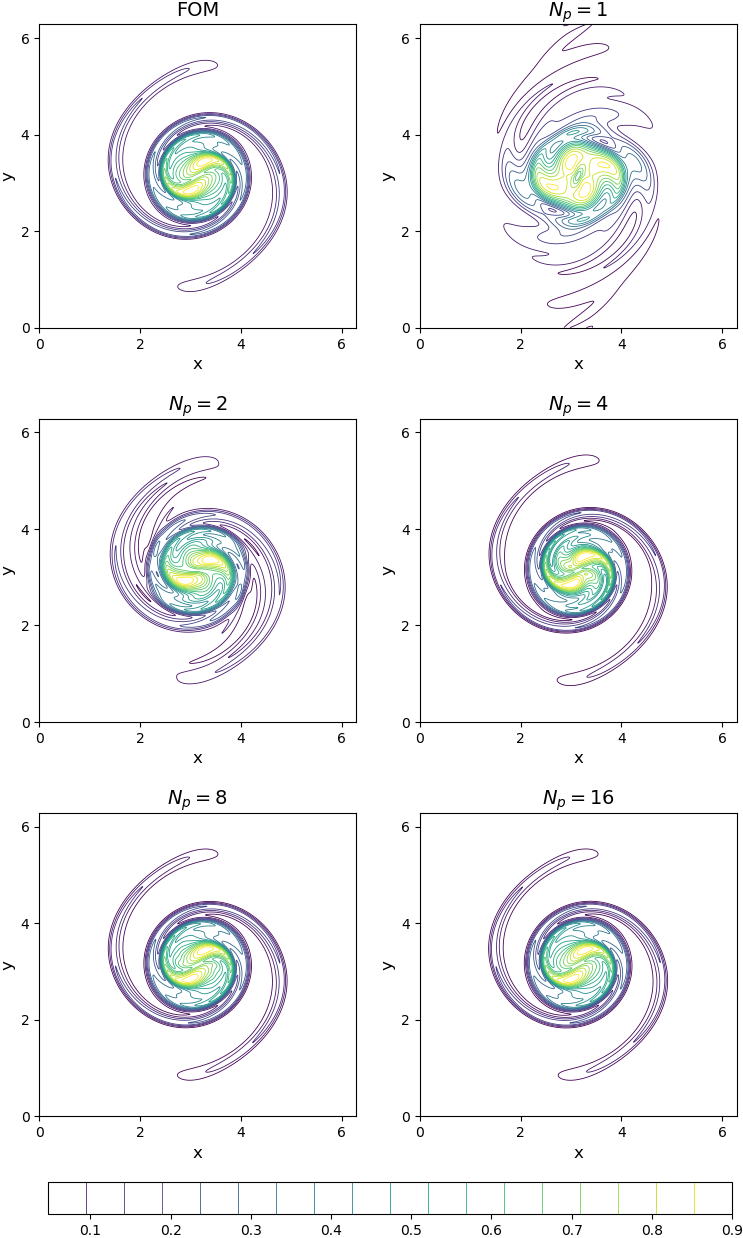}
	\caption{Final vorticity contours (i.e., at $t=40$) for vortex merger problem from PID-GP prediction using different number of intervals compared to FOM solution.}
	\label{fig:VM_GP}
\end{figure}

\begin{figure}[!ht]
	\centering
	\includegraphics[width=0.45\textwidth]{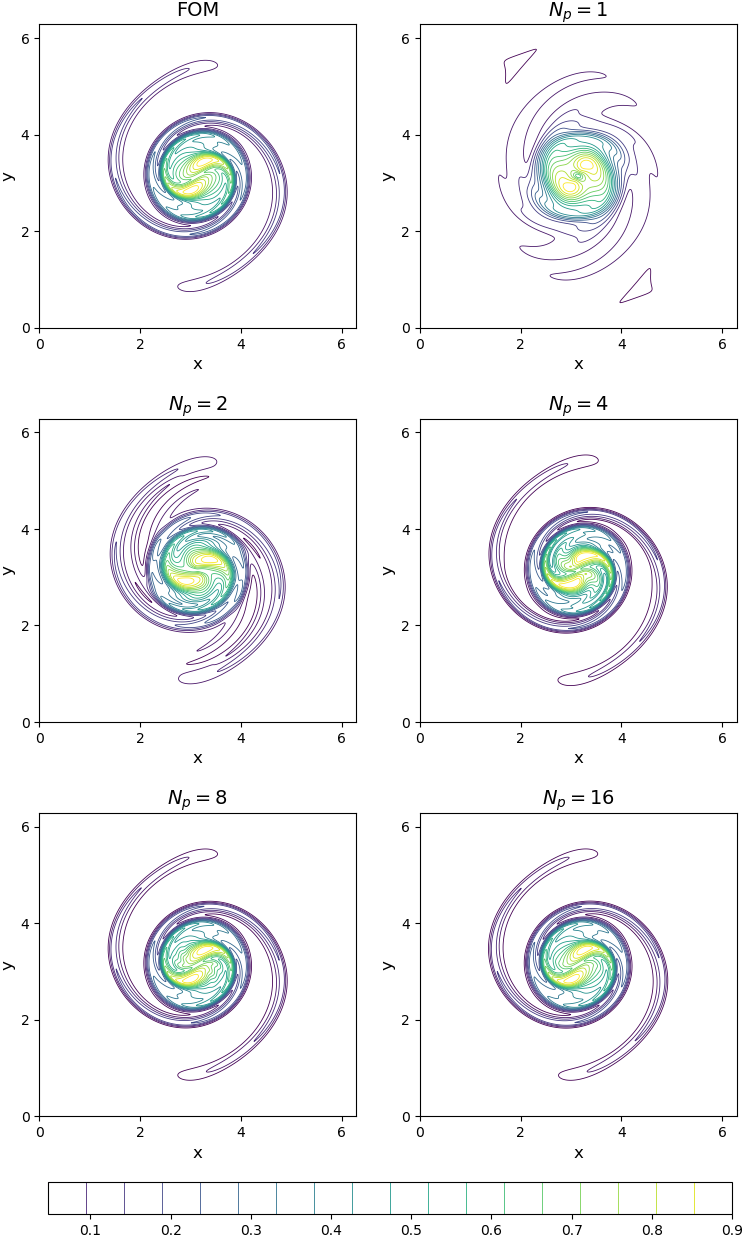}
	\caption{Final vorticity contours (i.e., at $t=40$) for vortex merger problem from PID-LSTM prediction using different number of intervals compared to FOM solution.}
	\label{fig:VM_LSTM}
\end{figure}

We compare the final field, characterizing the merging of two votices into a single vortex at the center of the 2D domain. Figure~\ref{fig:VM_POD} shows true projection, while Figure~\ref{fig:VM_GP} and~\ref{fig:VM_LSTM} illustrate the results obtained using the PID-GP framework and PID-LSTM framework, respectively. Similar to the 1D Burgers results, we note that increasing the number of intervals improves the results significantly. For example, if we look at the obtained field using a single interval, we will find a vorticity field that do not resemble the true one except for a very poor approximation of the merging phenomenon. However, the external field (away from the core) is very different from the true physical spiral motion. Moreover, Figure~\ref{fig:VM_GP} shows larger deformation in the field compared to both true projection and PID-LSTM framework. Again, this is due to the nonlinear interactions affecting the truncated ROM equations, requiring stabilization schemes to mitigate these effects.

\subsection{2D double shear layer problem} 
\label{sec:ShearRes}
\begin{figure}[!ht]
	\centering
	\includegraphics[width=0.45\textwidth]{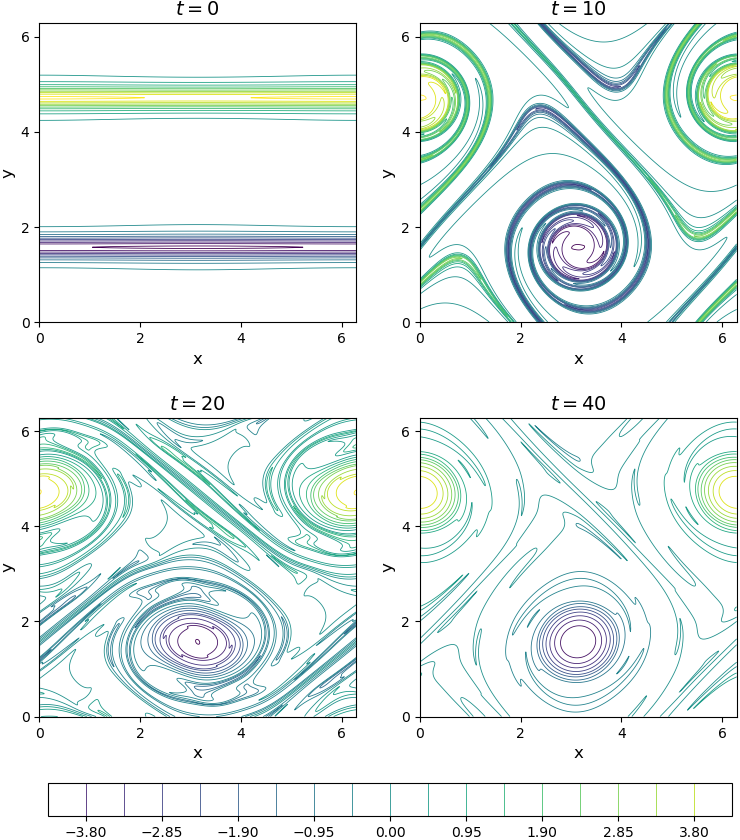}
	\caption{Vorticity field at different time instances for double shear layer problem using $1024^2$ grid and $\Delta t= 0.001$.}
	\label{fig:Dshear}
\end{figure}

Another application for 2D Navier-Stokes equations is the double shear layer problem, introduced by Bell et al \cite{bell1989second}. We consider a square domain of side length $2\pi$ with the following initial field \cite{san2012high},
\begin{equation}
    \omega(x, y, 0) = \begin{cases} \delta \cos(x) - \sigma \cosh^{-2}(\sigma[y-\dfrac{\pi}{2}])  & \mbox{if } y \le \pi, \\
                                    \delta \cos(x) + \sigma \cosh^{-2}(\sigma[\dfrac{3\pi}{2}-y]) & \mbox{if } y  > \pi. \end{cases}
\end{equation}
This field represents a horizontal shear layer of finite thickness (determined by $\delta$), perturbed by a small amplitude vertical velocity, where $\sigma$ determines the amplitude of this initial perturbation. In the present study, we adopt values of  $\delta = 0.05$ and $\sigma = 15/\pi$. Similar to the vortex merger setup, we use $\text{Re} = 10,000$ over a grid of $1024^2$ and $\Delta t = 0.001$. Indeed, the same numerical solver was used for both vortex merger and double shear layer problem, with only different initial conditions. The evolution of the double shear layer from time $t=0$ to $t=40$ is shown in Figure~\ref{fig:Dshear}, where the top and bottom shear layers evolve into a periodic array of large vortices,  and the layer between the rolls become thinner and thinner.

\begin{figure}[!ht]
	\centering
	\includegraphics[width=0.45\textwidth]{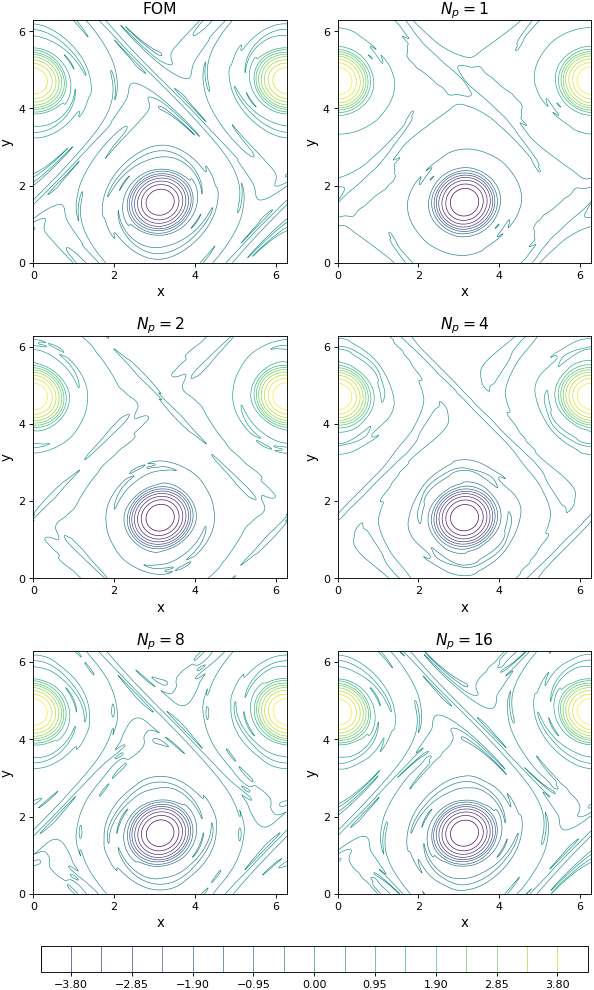}
	\caption{Final vorticity contours (i.e., at $t=40$) for double shear layer problem from true projection using different number of intervals compared to FOM solution.}
	\label{fig:DS_POD}
\end{figure}

\begin{figure}[!ht]
	\centering
	\includegraphics[width=0.45\textwidth]{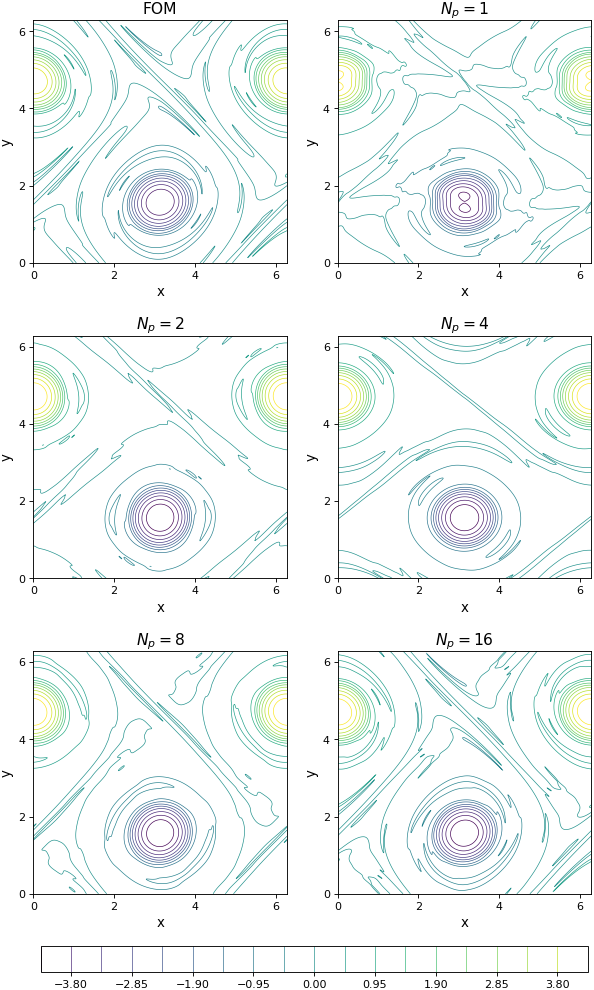}
	\caption{Final vorticity contours (i.e., at $t=40$) for double shear layer problem from PID-GP prediction using different number of intervals compared to  FOM solution.}
	\label{fig:DS_GP}
\end{figure}

\begin{figure}[!ht]
	\centering
	\includegraphics[width=0.45\textwidth]{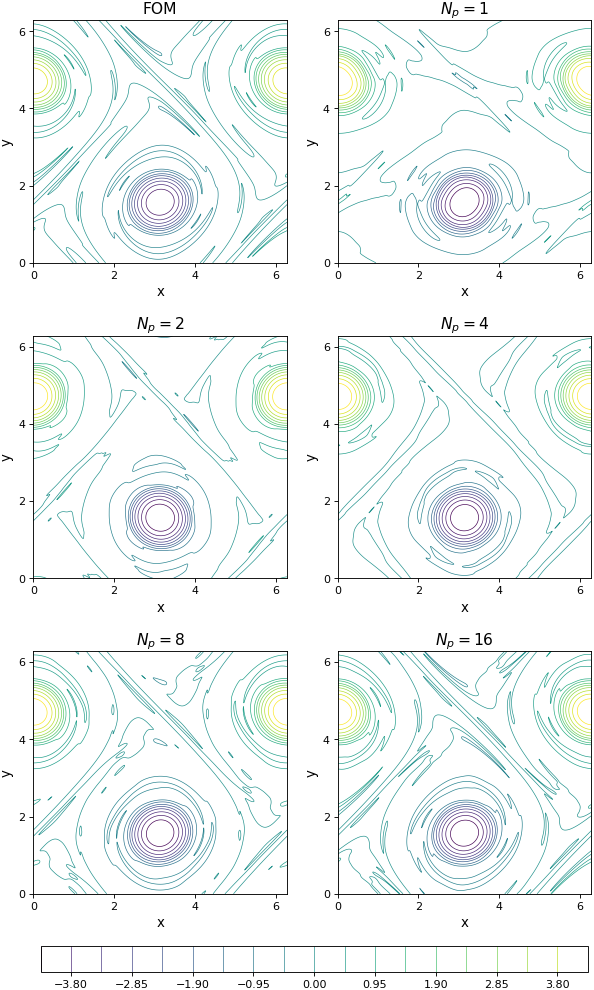}
	\caption{Final vorticity contours (i.e., at $t=40$) for double shear layer problem from PID-LSTM prediction using different number of intervals compared to  FOM solution.}
	\label{fig:DS_LSTM}
\end{figure}

Similar results are obtained for the double shear problem, as shown in Figures~\ref{fig:DS_POD},~\ref{fig:DS_GP}, and~\ref{fig:DS_LSTM}. As can be seen, significant details of the shear layers cannot be captured using a single global interval ($N_p=1$), even from the direct projection of the FOM field on the ROM space, Figure~\ref{fig:DS_POD}. The situation is even worse in PID-GP, where the vortex at the center of domain is deformed. Interestingly, the PID-LSTM performs much better than PID-GP, almost similar to the true projected fields.

\subsection{2D Boussinesq problem} 
\label{sec:BoussRes}
The two-dimensional Boussinesq problem is one-step more complex than the 2D Navier-Stokes equations, solving the energy equation along with the momentum equations. We consider a strong-shear flow exhibiting the Kelvin-Helmholtz instability, known as Marsigli flow or lock-exchange problem. The physical process in this flow problem explains how differences in  temperature/density can cause  currents  to  form  in  the ocean and seas. Basically, when fluids of two different densities meet, the higher density fluid slides below the lower density one. This is one of the primary mechanisms by which ocean currents are formed \cite{gill1982atmosphere}.

\begin{figure}[!ht]
	\centering
	\includegraphics[width=0.45\textwidth]{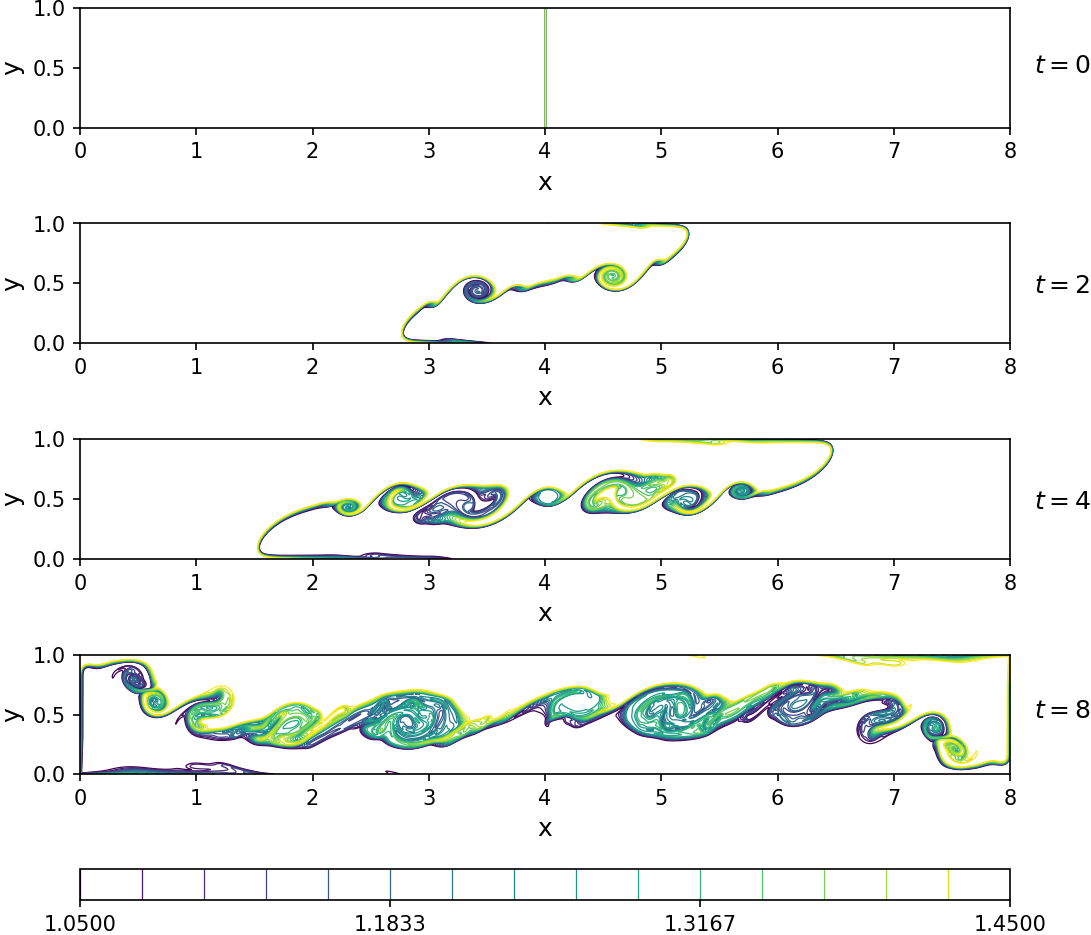}
	\caption{Temperature field at different time instances for 2D Boussinesq problem using $4096\times512$ grid and $\Delta t= 0.0005$.}
	\label{fig:Boussinesq}
\end{figure}

We consider two fluids of different temperatures, in a rectangular domain $(x,y) \in [0,8] \times [0,1]$. A vertical barrier divides the domain at $x=4$, keeping the temperature, $\theta$, of the left half at $1.5$ and temperature of the right half at $1$. Initially, the flow is at rest (i.e., $\omega(x,y,0) = \psi(x,y,0) = 0$, with uniform temperatures at the right and left regions (i.e., $\theta(x,y,0) = 1.5 \ \forall \ x\in [0,4]$ and $\theta(x,y,0) = 1 \ \forall \ x\in (4,8]$). No-slip boundary conditions are assumed for flow field, and adiabatic boundary conditions are prescribed for temperature field. Reynolds number of $\text{Re} = 10^4$, Richardson number of $\text{Ri} = 4$, and Prandtl number of $\text{Pr}=1$ are set in Equations~\ref{eq:Bouss1}-\ref{eq:Bouss2}. A Cartesian grid of $4096\times512$, and a timestep of $\Delta t=5\times10^{-4}$ are used for the FOM simulations. The evolution of the temperature field is shown in Figure~\ref{fig:Boussinesq} at $t=0,2,4,8$. At time zero, the barrier is removed instantaneously triggering the lock-exchange problem. Due to the temperature difference (causing density difference), buoyancy forces start to emerge. The higher density fluid (on the right) slides below the lower density fluid (on the left) causing an undercurrent flow moving from right to left. Conversely, an upper current flow moves from left to right, causing a strong shear layer between the countercurrent flows. As a result, vortex sheets are produced, exhibiting the Kelvin-Helmholtz instability. This problem is challenging, even for DNS simulations \cite{liu2003fourth}, making it a good benchmark for POD/PID comparison. 

\begin{figure}[!ht]
	\centering
	\includegraphics[width=0.45\textwidth]{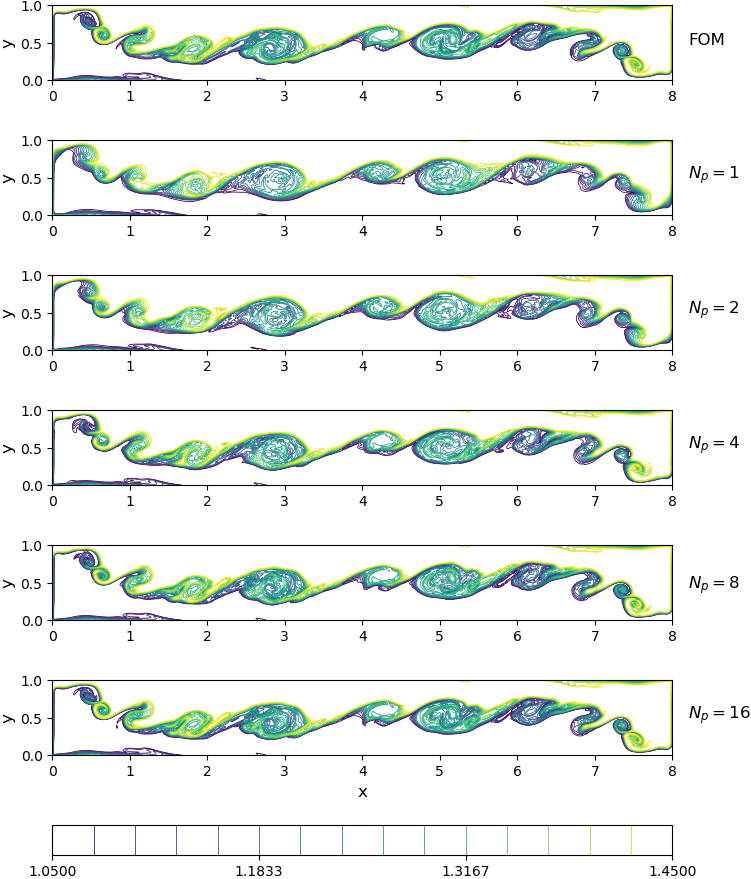}
	\caption{Final contours (i.e., at $t=8$) for Boussinesq problem from true projection using different number of intervals compared to FOM solution.}
	\label{fig:BQ_POD}
\end{figure}

\begin{figure}[!ht]
	\centering
	\includegraphics[width=0.45\textwidth]{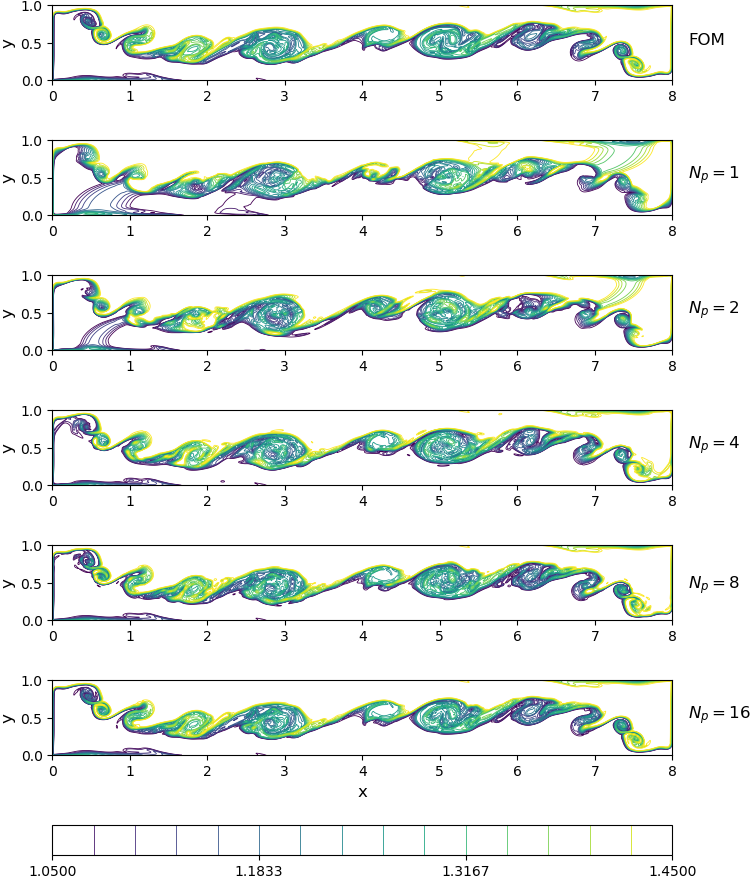}
	\caption{Final contours (i.e., at $t=8$) for Boussinesq problem from PID-GP prediction using different number of intervals compared to FOM solution.}
	\label{fig:BQ_GP}
\end{figure}

\begin{figure}[!ht]
	\centering
	\includegraphics[width=0.45\textwidth]{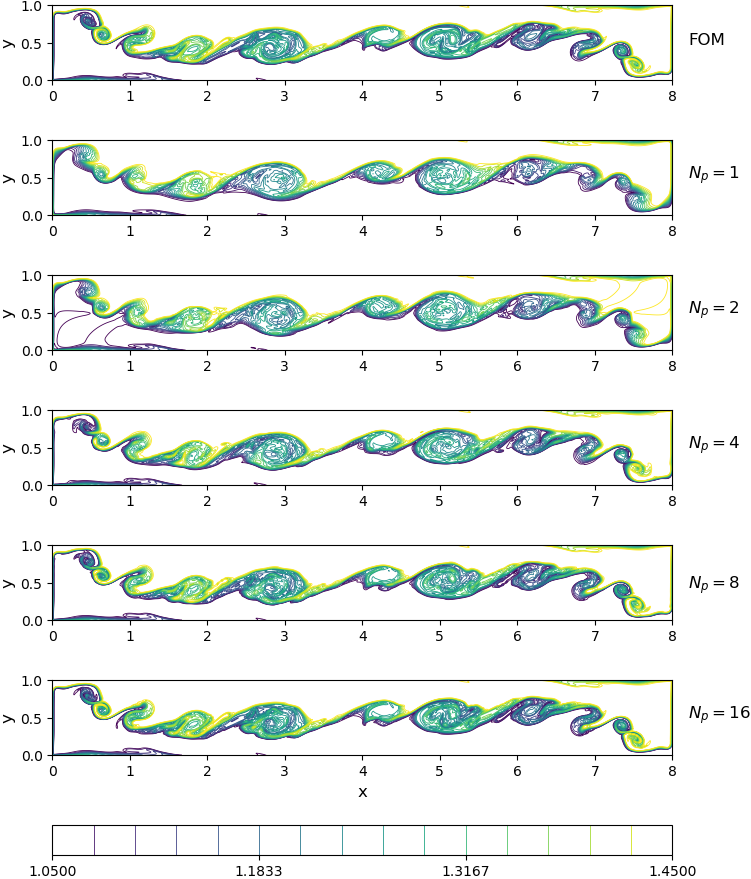}
	\caption{Final contours (i.e., at $t=8$) for Boussinesq problem from PID-LSTM prediction using different number of intervals compared to FOM solution.}
	\label{fig:BQ_LSTM}
\end{figure}

In Figure~\ref{fig:BQ_POD}, we show the true projection of the final temperature field on the POD/PID space using different number of intervals. Although the overall structure is represented nicely using a single interval (corresponding to standard POD), the small-scale structures are not captured. If we investigate the contour lines carefully, we can see that standard POD smooth-out the field. As the number of intervals is increased, more details can be captured using local basis functions. Also, we compare the final temperature field predictions from standard Galerkin projection and LSTM frameworks. Contour plots for final temperature field are shown in Figures~\ref{fig:BQ_GP}-\ref{fig:BQ_LSTM} at different number of intervals, $N_p$. Similar to previous cases, PID-GP is adding more deformation to the results and instabilities are amplified. This is due to the fact that the eigenvalues of this flow problem are decaying slowly, especially for such high $\text{Re}$ used in current study. In the present study we used just 6 modes (i.e., $R=6$), corresponding to $RIC$ of only $60.97 \%$ for vorticity and $88.27$ for temperature fields using one interval. Due to the dependence of Galerkin projection on the governing equations, the resulting ROMs strongly couples temperature and vorticity. Therefore, inaccuracies in vorticity predictions often affects temperature predictions and vice versa since they are coupled with each others in the ROM-GP equations. So even if 6 modes with $N_p=4$ can capture more than $90\%$ of the variance in temperature field, the results will be affected by the low energy captured from the voriticty fields (less than $70\%$). On the other hand, LSTM predictions do not have this dependency, which enables us to use the most accurate datasets and fields with the maximum reducibility without being affected by other irrelevant fields.

\begin{figure}[!ht]
	\centering
	\includegraphics[width=0.45\textwidth]{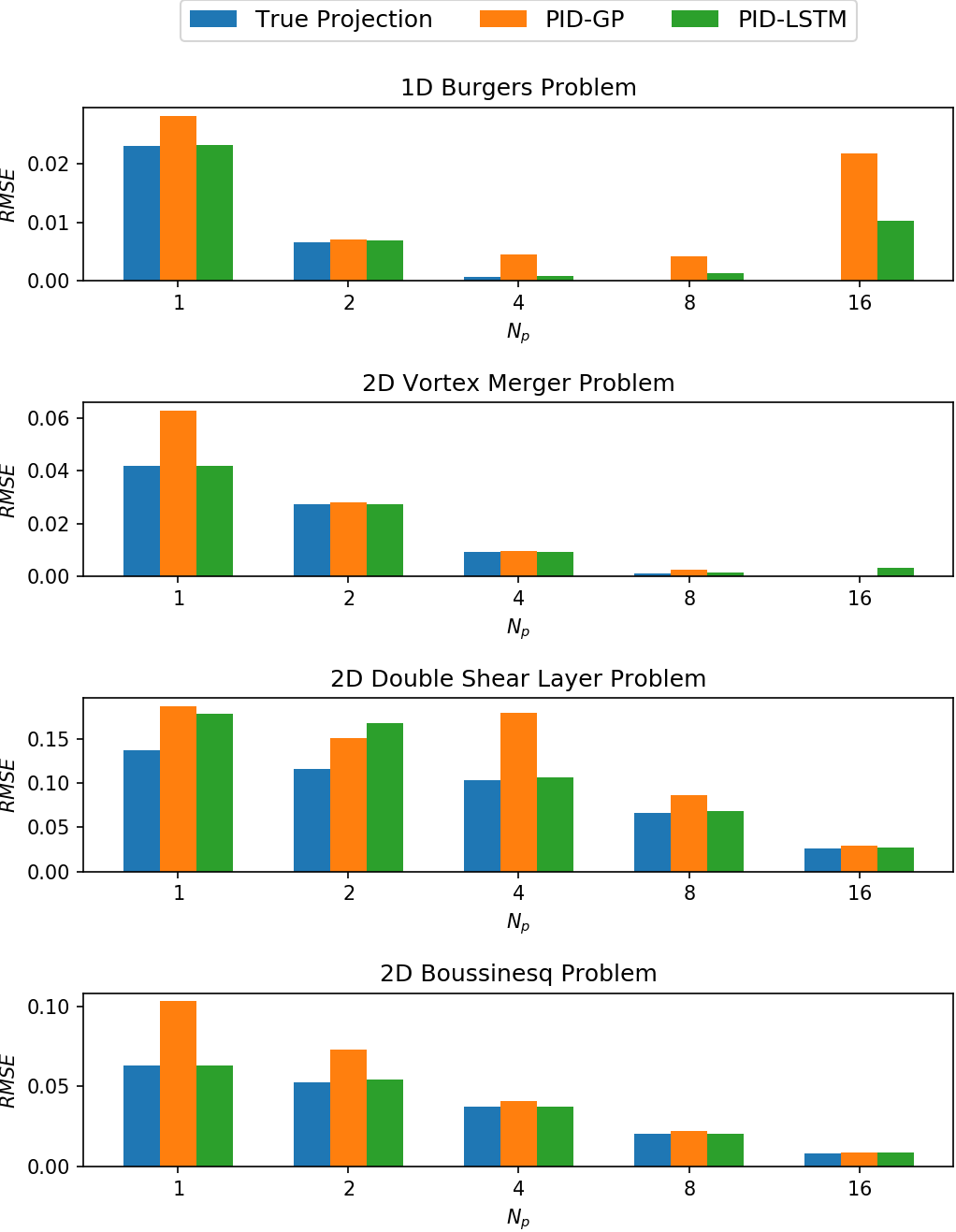}
	\caption{A bar chart for $RMSE$ of predicted field at final time.}
	\label{fig:RMSE}
\end{figure}

Finally, in order to quantify the results in a more quantitative way, we use the root mean square error ($RMSE$) as an error measure, defined as
\begin{equation}
    RMSE(t) = \sqrt{\dfrac{1}{N} \sum_{i=1}^{N} \big(u^{FOM}(\mathbf{x},t) - u(\mathbf{x},t) \big)^2 },
\end{equation}
where $N$ is the spatial resolution, as defined in Section~\ref{sec:pod}. For 1D cases, it is simply $N_x$, and in 2D cases it is ($N_x\times N_y$). We calculated the RMSE at the final field (i.e. at $t=T$) using different approaches. Results are given in Table~\ref{tab:RMSE} and illustrated graphically using a bar chart in Figure~\ref{fig:RMSE} confirming our earlier findings about the accuracy gain due to time decomposition as well as PID-LSTM being superior to PID-GP framework.

\begin{table}[ht!]
	\centering
	\caption{$RMSE$ for predicted field at final time from true projection, PID-GP framework, and PID-LSTM framework compared to FOM results.}
	\begin{tabular}[t]{l  l  l  l} 
		\hline\noalign{\smallskip}
		$N_p$ \quad \quad \quad &True Projection \quad \quad \quad &PID-GP \quad \quad \quad &PID-LSTM   \\ 
		\noalign{\smallskip}\hline \noalign{\smallskip}
		\multicolumn{4}{l}{\emph{\underline{1D Burgers}}}\smallskip \\
		$1$ & \ $2.31\text{E}-2$ & $2.82\text{E}-2$ & \ $2.32\text{E}-2$ \\
        $2$ & \ $6.62\text{E}-3$ & $7.02\text{E}-3$ & \ $6.88\text{E}-3$ \\
        $4$ & \ $7.24\text{E}-4$ & $4.50\text{E}-3$ & \ $7.93\text{E}-4$ \\
        $8$ & \ $2.16\text{E}-5$ & $4.23\text{E}-3$ & \ $1.39\text{E}-3$ \\
        $16$& \ $2.19\text{E}-7$ & $2.18\text{E}-2$ & \ $1.03\text{E}-2$  \medskip \\ 
		\multicolumn{4}{l}{\emph{\underline{2D Vortex Meger}}}\smallskip \\
		$1$ & \ $4.18\text{E}-2$ & $6.27\text{E}-2$ & \ $4.20\text{E}-2$ \\
        $2$ & \ $2.74\text{E}-2$ & $2.79\text{E}-2$ & \ $2.75\text{E}-2$ \\
        $4$ & \ $9.08\text{E}-3$ & $9.57\text{E}-3$ & \ $9.14\text{E}-3$ \\
        $8$ & \ $1.09\text{E}-3$ & $2.35\text{E}-3$ & \ $1.61\text{E}-3$ \\
        $16$& \ $4.80\text{E}-5$ & $1.65\text{E}-4$ & \ $3.29\text{E}-3$   \medskip \\ 
		\multicolumn{4}{l}{\emph{\underline{2D Double Shear Layer}}}\smallskip \\
        $1$ & \ $1.37\text{E}-1$ & $1.87\text{E}-1$ & \ $1.78\text{E}-1$ \\
        $2$ & \ $1.16\text{E}-1$ & $1.51\text{E}-1$ & \ $1.68\text{E}-1$ \\
        $4$ & \ $1.03\text{E}-1$ & $1.79\text{E}-1$ & \ $1.06\text{E}-1$ \\
        $8$ & \ $6.63\text{E}-2$ & $8.69\text{E}-2$ & \ $6.83\text{E}-2$ \\
        $16$& \ $2.56\text{E}-2$ & $2.97\text{E}-2$ & \ $2.67\text{E}-2$  \medskip \\ 
		\multicolumn{4}{l}{\emph{\underline{2D Boussinesq}}}\smallskip \\
		$1$ & \ $	6.30\text{E}-2$ & $1.03\text{E}-1$ & \ $6.30\text{E}-2$ \\
        $2$ & \ $5.28\text{E}-2$ & $7.28\text{E}-2$ & \ $5.45\text{E}-2$ \\
        $4$ & \ $3.73\text{E}-2$ & $4.11\text{E}-2$ & \ $3.73\text{E}-2$ \\
        $8$ & \ $2.03\text{E}-2$ & $2.22\text{E}-2$ & \ $2.03\text{E}-2$ \\
        $16$& \ $8.24\text{E}-3$ & $8.94\text{E}-3$ & \ $8.67\text{E}-3$ \\
		\hline
	\end{tabular}
	\label{tab:RMSE}
\end{table}

\begin{figure}[!ht]
	\centering
	\includegraphics[width=0.45\textwidth]{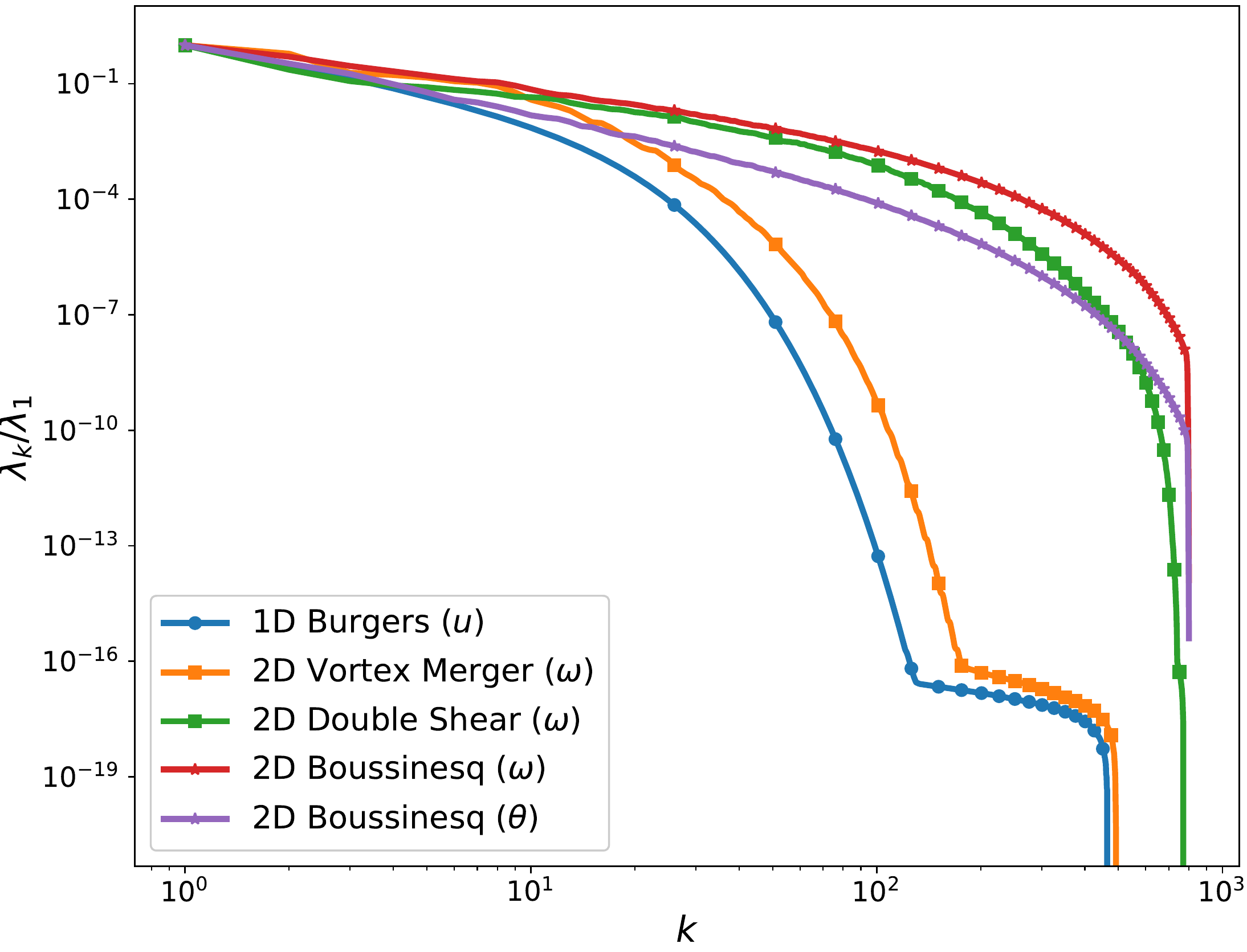}
	\caption{Eigenspectrum plot for the tested cases (normalized with respect to the first (largest) eigenvalue).}
	\label{fig:eigs}
\end{figure}

An important observation from Table~\ref{tab:RMSE} and Figure~\ref{fig:RMSE} is the significant increase in $RMSE$ for Burgers problem at $N_p=16$. To express this behavior, we plotted the eigenspectrum for these four problems as shown in Figure~\ref{fig:eigs}. Interestingly, we can see that the decay of eigenvalues for 1D Burgers case is faster than other cases, while the decay of eigenvalues in 2D Boussinesq problem is the slowest. This implies that the higher the decay rate is, the fewer intervals are required. That is for 1D Burgers case, 4 or 8 subintervals are more than enough to capture local dynamics, and further increase in the number of partitions leads to an increase in $RMSE$. This might be caused by the degradation of field quality due to successive reconstructions/projections at the interface. After a finite number of these interface treatments, the accuracy gain due to localization is surpassed by that successive degradation. This can be mitigated by applying closure and/or regularization techniques at the interface to enhance the reconstructed field and account for truncated modes before transferring into the subsequent manifold. On the other hand, for more challenging problems when the rate of decay is slow (such as 2D Double Shear and 2D Boussinesq cases), more localization helps to increase predictive accuracy of PID. This implies that for a higher Kolmogorov $n$-width barrier (i.e., lower decay rate), a larger number of subintervals $N_p$ is required and the PID approach can offer a viable solution in such situations.

\section{Conclusions}
\label{sec:conc}
In the current study, we presented a fully data-driven non-intrusive framework for convection-dominated problems. Most model reduction techniques depend on the ergodicity hypothesis which implies that any ensemble of realizations would carry the average statistical properties of the entire process. Hoewever in convective flow problems (like those investigated in present study), the convective mechanisms are more dominant than diffusive ones. The ergodicity hypothesis is therefore violated, making the application of standard model reduction algorithms infeasible. Moreover, the Kolomogorov barrier constraints the reducibility of such systems. We address these issues by employing a splitting technique, based on principal interval decomposition. We divided our time domain into a set of equidistant partitions, and applied the POD locally in each of them as our compression approach. For system's dynamics (encapsulated in temporal coefficients), we trained corresponding LSTM models at each zone equipped with consisted interface conditions. PID-LSTM results were compared with standard Galerkin projection framework. It was found that PID-GP provided less accurate results than PID-LSTM. This was particularly evident in problems where multiple fields are coupled, such 2D Boussinesq case. On the other side, PID-LSTM enables us to separate the quantity of interest and deploy our prediction on the most relevant problem-specific ones. Therefore, for example, the temperature field can be inferred without explicitly constructing the vorticity fields because of the non-intrusive nature of the predictive modeling framework. We also observed that for convection-dominated problems the optimal number of intervals is dependent on the decay rate of Kolomogorov $n$-width. Hence, we suggest that adaptive and automated partitioning or clustering techniques would be useful in this context. Finally, since the most expensive stages of PID-LSTM (decomposition and training) are performed offline, it is capable of providing near real-time responses during the online stage. This can serve as a key enabler for developing digital twin technologies, a topic that we plan to cover more in detail in the future.

\section*{Acknowledgements}
This material is based upon work supported by the U.S. Department of Energy, Office of Science, Office of Advanced Scientific Computing Research under Award Number DE-SC0019290. 
OS gratefully acknowledges their support. Disclaimer: This report was prepared as an account of work sponsored by an agency of the United States Government. Neither the United States Government nor any agency thereof, nor any of their employees, makes any warranty, express or implied, or assumes any legal liability or responsibility for the accuracy, completeness, or usefulness of any information, apparatus, product, or process disclosed, or represents that its use would not infringe privately owned rights. Reference herein to any specific commercial product, process, or service by trade name, trademark, manufacturer, or otherwise does not necessarily constitute or imply its endorsement, recommendation, or favoring by the United States Government or any agency thereof. The views and opinions of authors expressed herein do not necessarily state or reflect those of the United States Government or any agency thereof.

\section*{Appendix: Basis Functions}
Here, we visualize the constructed basis functions of POD and PID approaches using different number of intervals. Specifically, we present the first 4 modes of POD application on the whole time interval. This illustrates the deformation of obtained modes by the severely varying systems states with time. Also, we present the first mode computed locally in the first and last intervals (i.e., $\phi_1^{(1)}$ and $\phi_1^{(N_p)}$, respectively).

\subsection*{Burgers problem}
\begin{figure}[!ht]
	\centering
	\includegraphics[width=0.45\textwidth]{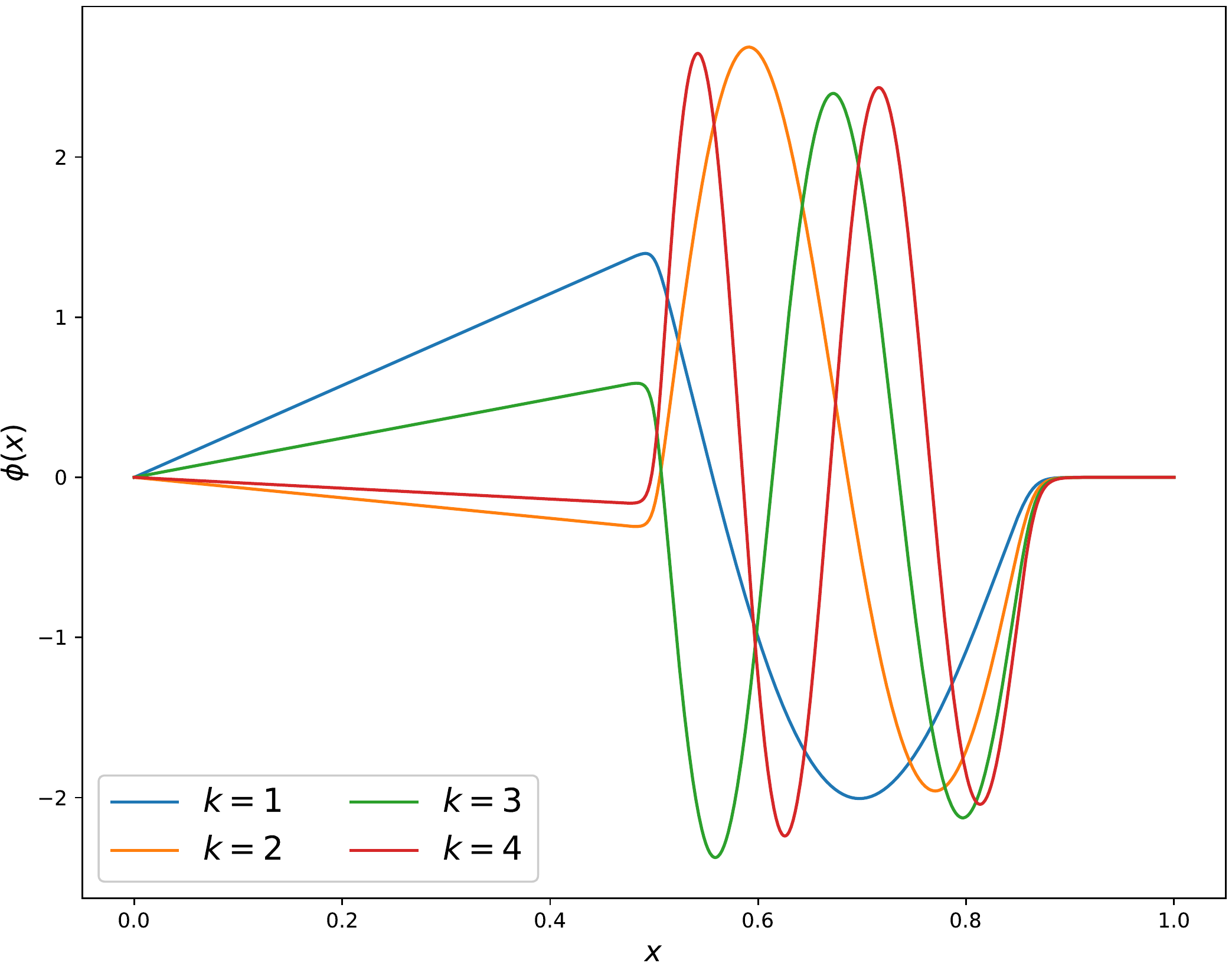}
	\caption{The first 4 global basis functions from POD application over the whole time domain (i.e., for $0\le t \le2$) for Burgers problem.}
	\label{fig:brg_basis0}
\end{figure}

Figure~\ref{fig:brg_basis0} shows the first four global functions calculated from the classical POD approach over the whole time interval. It can be observed that the shock is smoothed-out because it is moving with time. As a result, none of these modes resemble the actual state of the flow and not much information about the location and characteristics of these shocks can be inferred from these global modes.

\begin{figure}[!ht]
	\centering
	\includegraphics[width=0.45\textwidth]{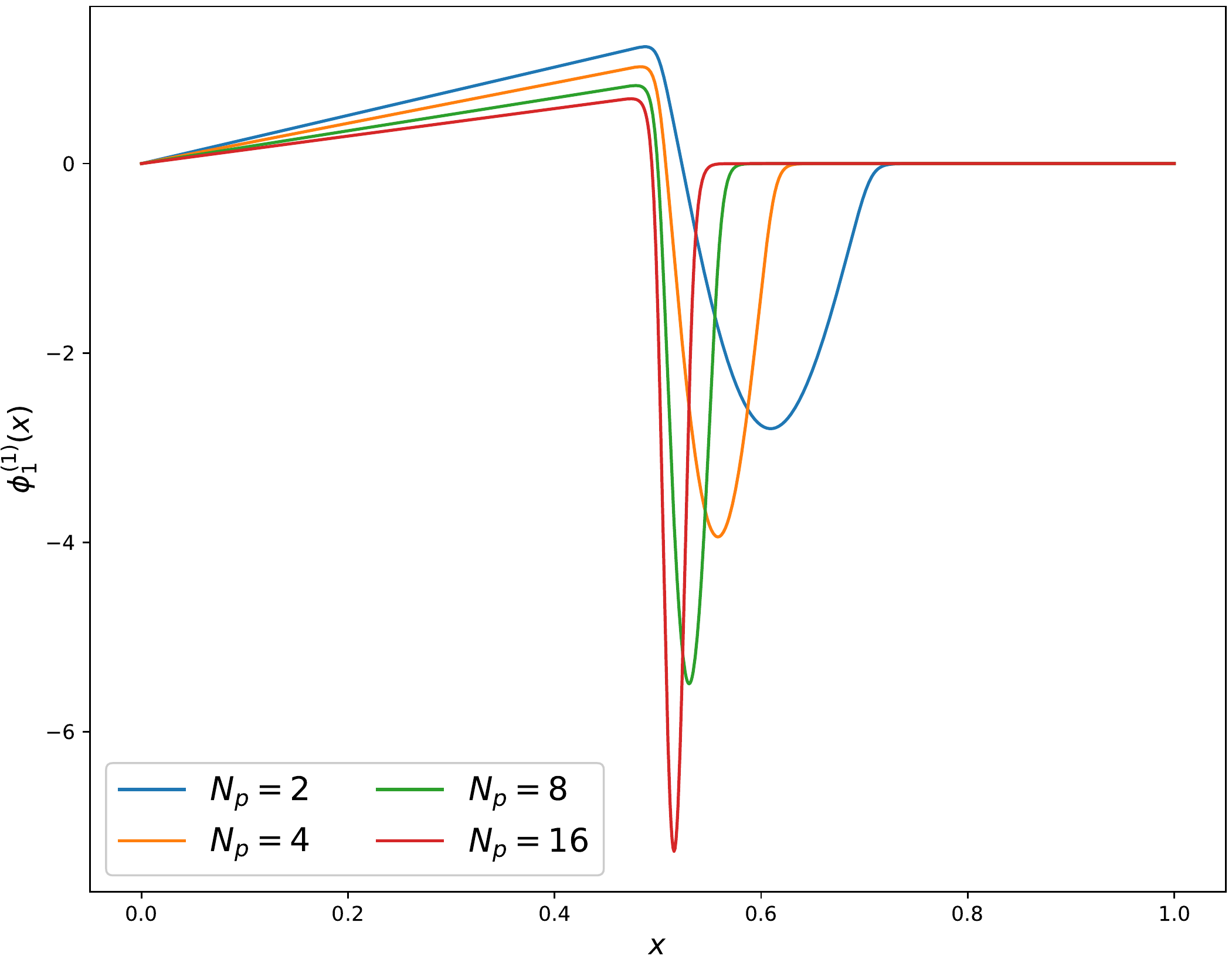}
	\caption{The first local basis function from PID application over the first subinterval (i.e., for $0\le t \le t_{\kappa^{(1)}}$) for Burgers problem using different number of intervals.}
	\label{fig:brg_basis1}
\end{figure}

\begin{figure}[!ht]
	\centering
	\includegraphics[width=0.45\textwidth]{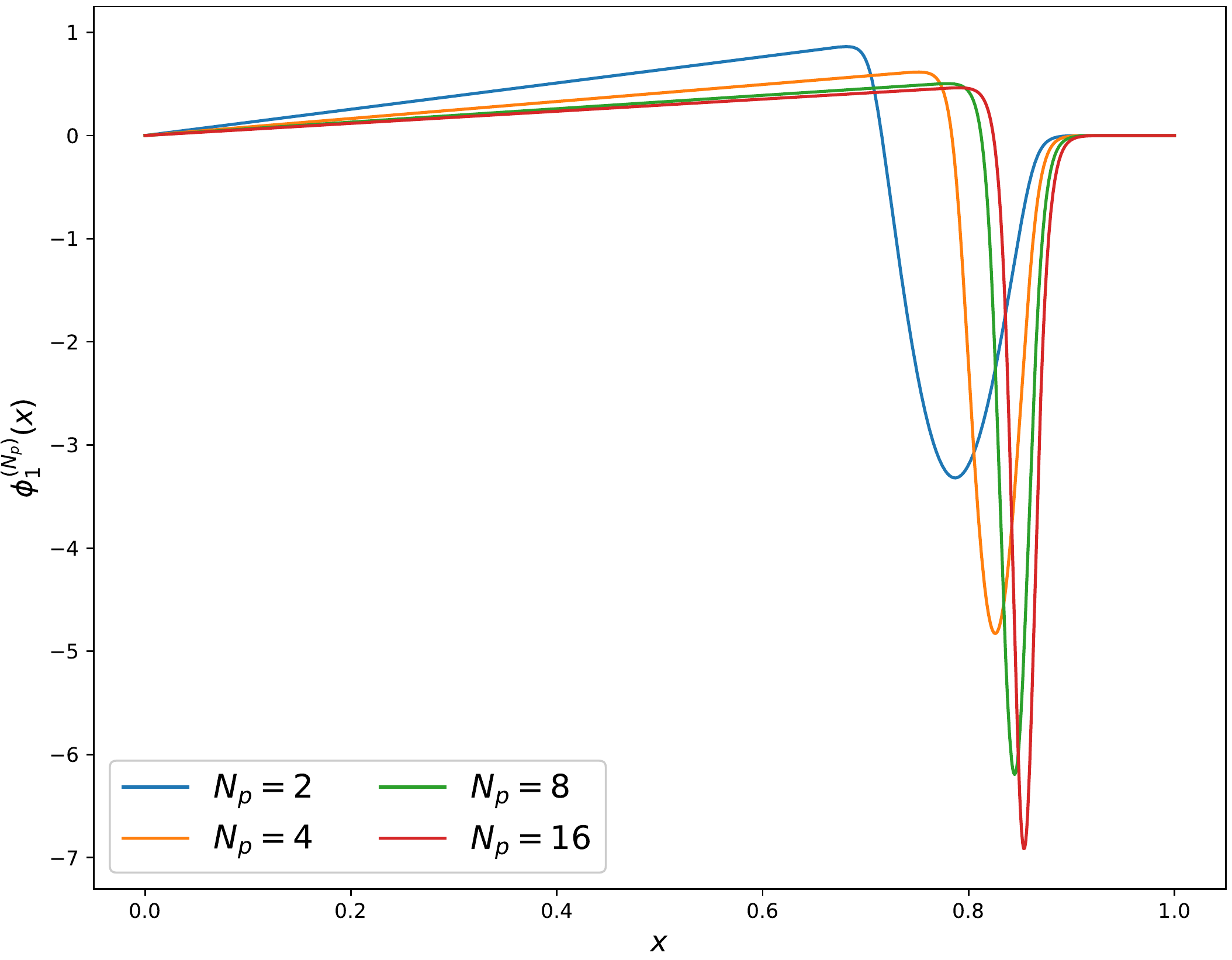}
	\caption{The first local basis function from PID application over the last subinterval (i.e., for $t_{\kappa^{(N_p-1)}}\le t \le 2 $) for Burgers problem using different number of intervals.}
	\label{fig:brg_basis2}
\end{figure}

On the other hand, the application of PID, results in local modes which give much better information about the shock characteristics. For example, the first mode in the first subinterval provides more accurate information about the initial shock location as emphasized in Figure~\ref{fig:brg_basis1}. As the number of intervals increases, the detection of the shock-wave is improved. Similar results are obtained in Figure~\ref{fig:brg_basis2}, where the shock wave at the final time is captured.

\subsection*{Vortex merger problem}
\begin{figure}[!ht]
	\centering
	\includegraphics[width=0.45\textwidth]{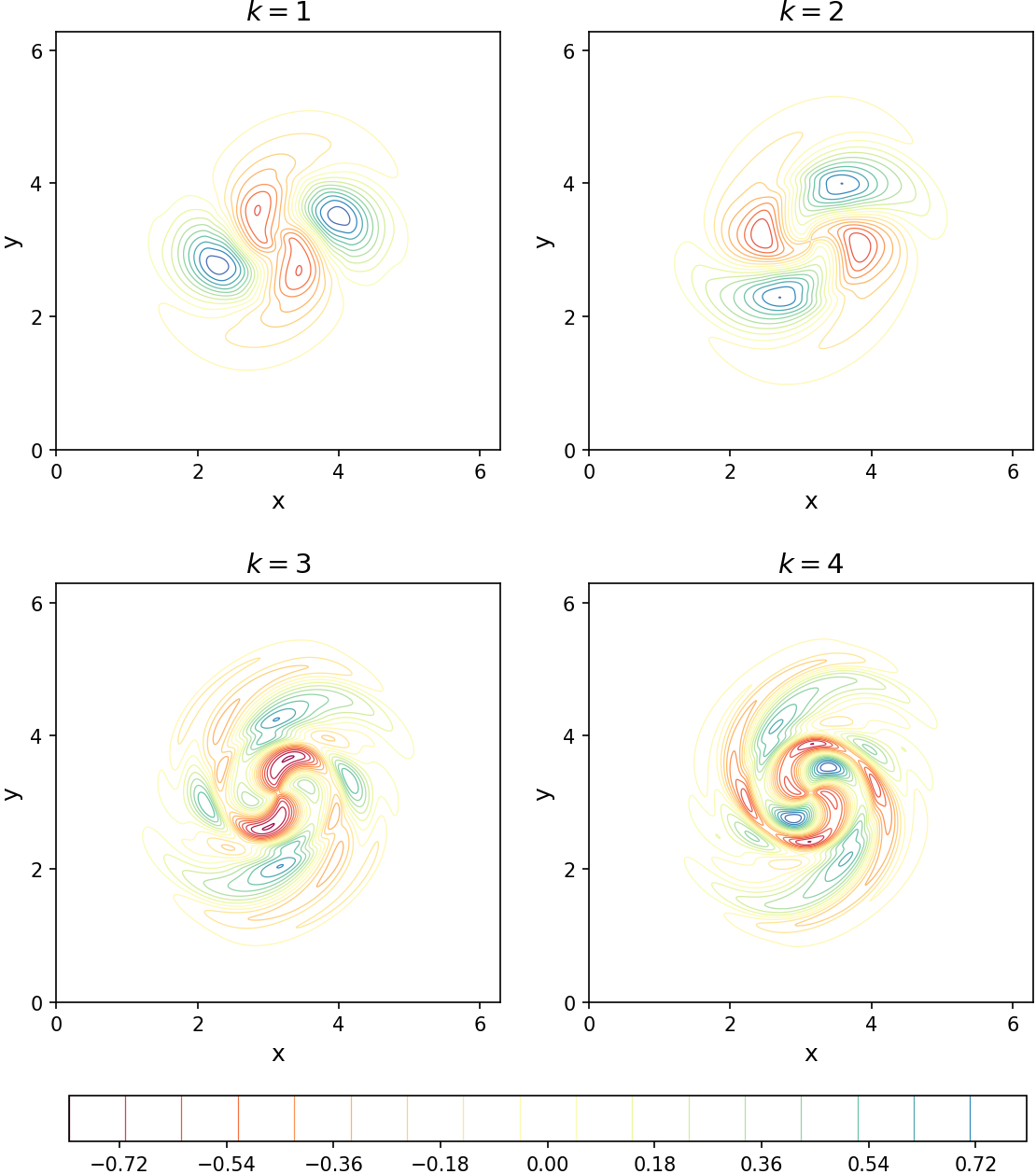}
	\caption{The first 4 global basis functions for vorticity field from POD application over the whole time domain (i.e., for $0\le t \le40$) for vortex merger problem.}
	\label{fig:VM_basis0}
\end{figure}

\begin{figure}[!ht]
	\centering
	\includegraphics[width=0.45\textwidth]{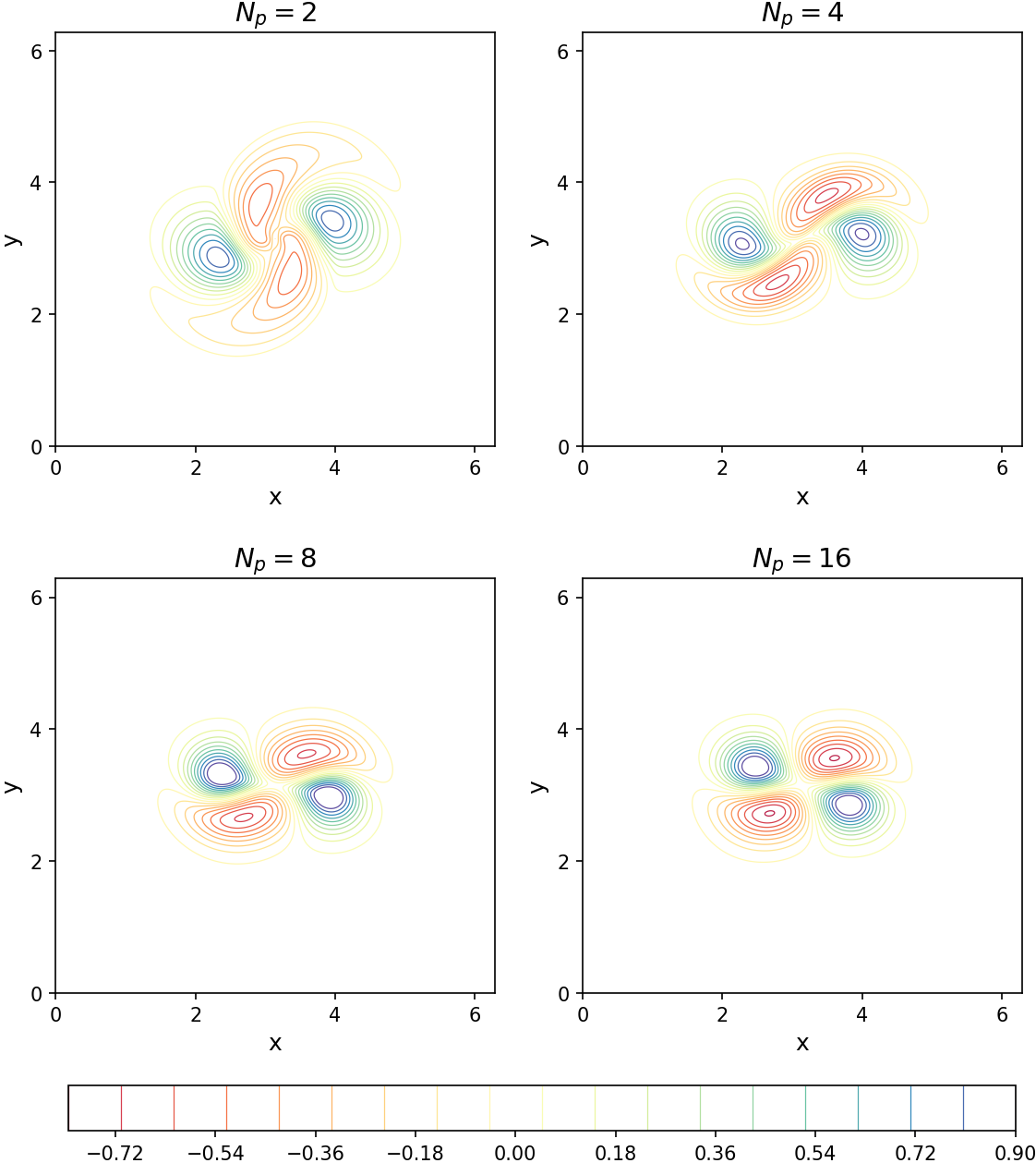}
	\caption{The first local basis function for vorticity field from PID application over the first subinterval (i.e., for $0\le t \le t_{\kappa^{(1)}}$) for vortex merger problem using different number of intervals.}
	\label{fig:VM_basis1}
\end{figure}

\begin{figure}[!ht]
	\centering
	\includegraphics[width=0.45\textwidth]{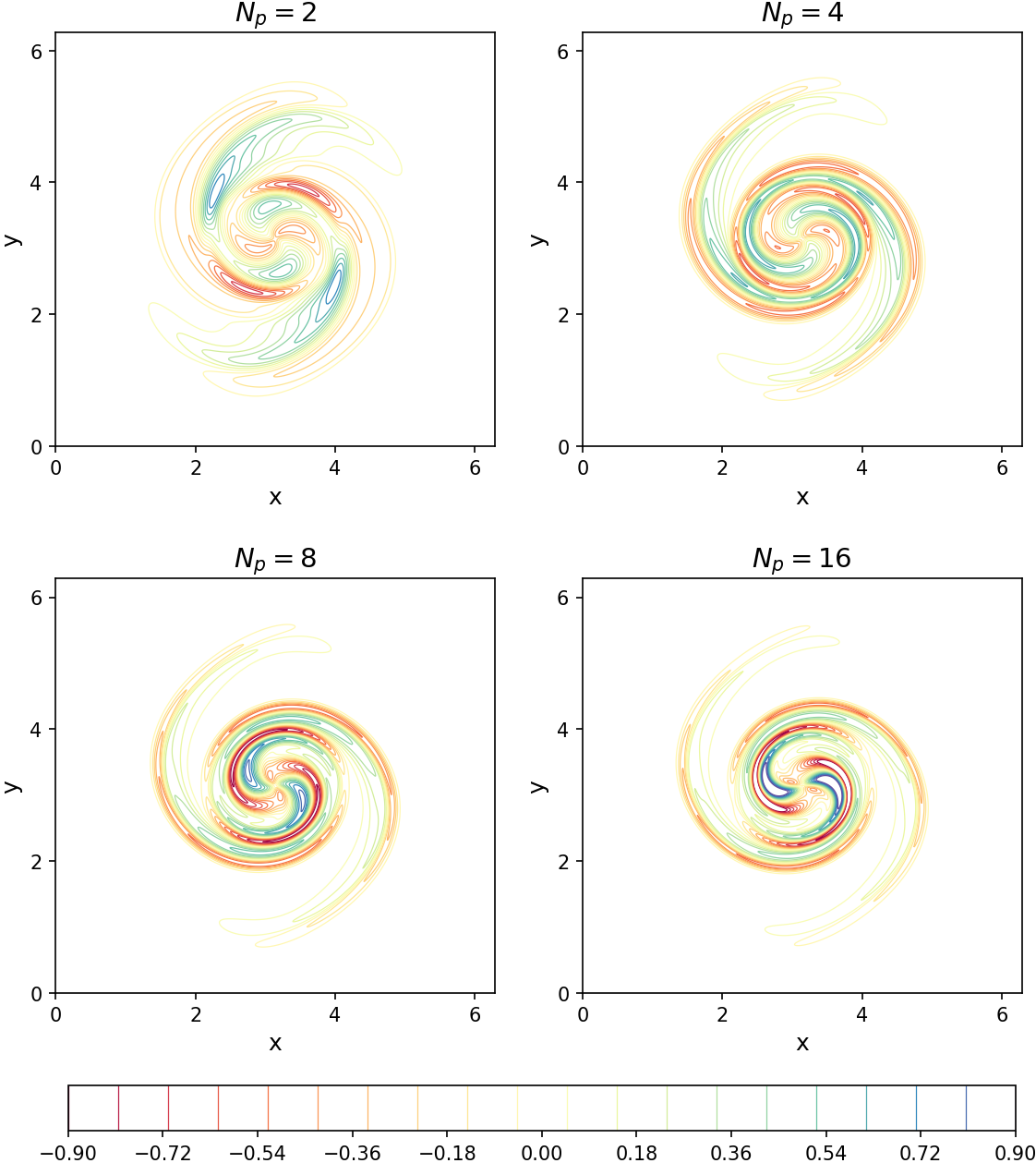}
	\caption{The first local basis function for vorticity field from PID application over the first subinterval (i.e., for $t_{\kappa^{(N_p-1)}}\le t \le 40 $) for vortex merger problem using different number of intervals.}
	\label{fig:VM_basis2}
\end{figure}

The contour plots for the global POD modes for vortex merger problem is shown in Figure~\ref{fig:VM_basis0}. We can easily observe the modal deformation in such a way that the obtained modes give just an overview of the merging process without much details about the growth of the two vortices. On the other hand, we can get more insights about the initial and final stages of the merging process from Figures~\ref{fig:VM_basis1}-\ref{fig:VM_basis2}, respectively.

\subsection*{Double shear layer problem}
\begin{figure}[!ht]
	\centering
	\includegraphics[width=0.45\textwidth]{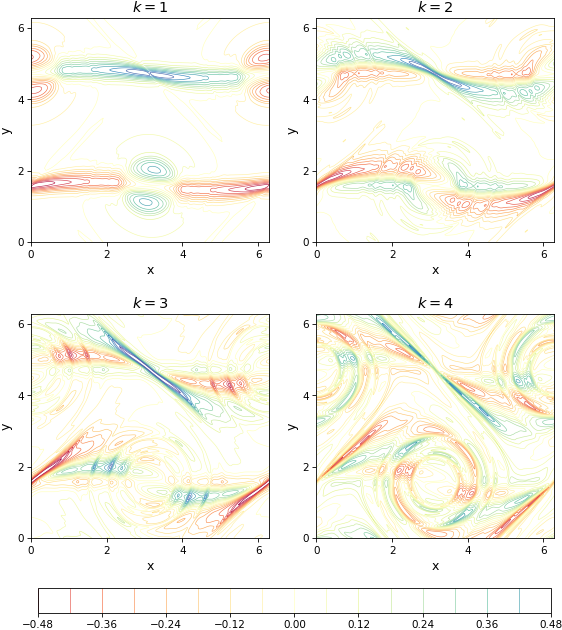}
	\caption{The first 4 global basis functions for vorticity field from POD application over the whole time domain (i.e., for $0\le t \le40$) for double shear layer problem.}
	\label{fig:DS_basis0}
\end{figure}

\begin{figure}[!ht]
	\centering
	\includegraphics[width=0.45\textwidth]{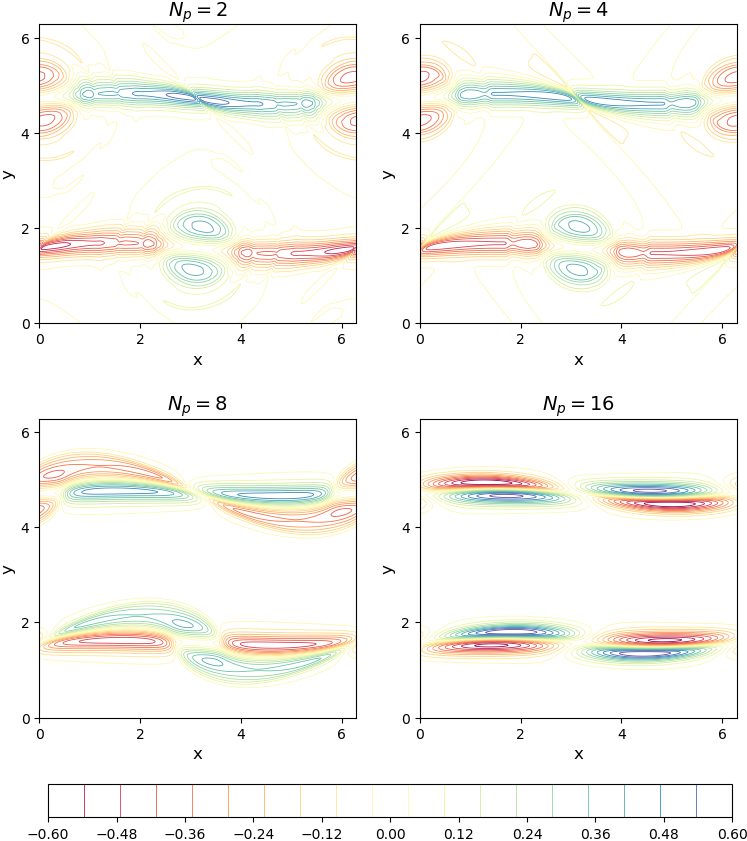}
	\caption{The first local basis function for vorticity field from PID application over the first subinterval (i.e., for $0\le t \le t_{\kappa^{(1)}}$) for double shear layer problem using different number of intervals.}
	\label{fig:DS_basis1}
\end{figure}

\begin{figure}[!ht]
	\centering
	\includegraphics[width=0.45\textwidth]{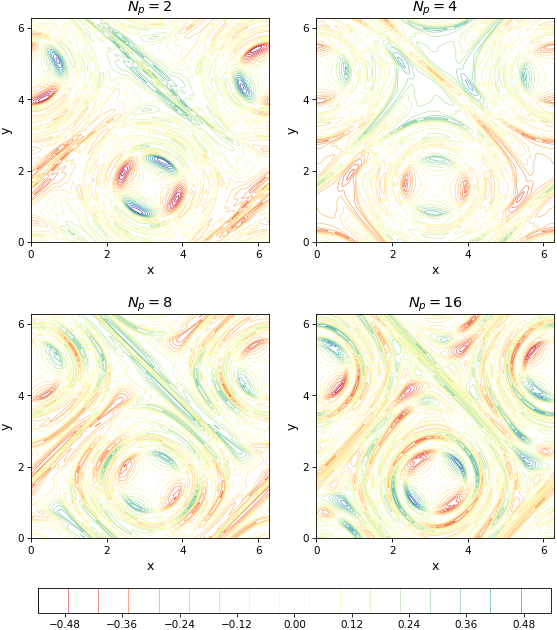}
	\caption{The first local basis function for vorticity field from PID application over the first subinterval (i.e., for $t_{\kappa^{(N_p-1)}}\le t \le 40 $) for double shear layer problem using different number of intervals.}
	\label{fig:DS_basis2}
\end{figure}

Similar results are obtained for the double shear layer problem, where the global deformed POD modes are shown in Figure~\ref{fig:DS_basis0}. It is clear that the final field structures are captured by higher modes, where the first mode barely carries any information about the final field. Therefore, more modes need to be retained in the ROM. On the other hand, the local basis function for the first subinterval is similar to the initial field as shown in Figure~\ref{fig:DS_basis1}, while the first basis function in the last subinterval captures the main dynamics of the final field as depicted in Figure~\ref{fig:DS_basis2}.

\subsection*{Boussinesq problem}

\begin{figure}[!ht]
	\centering
	\includegraphics[width=0.45\textwidth]{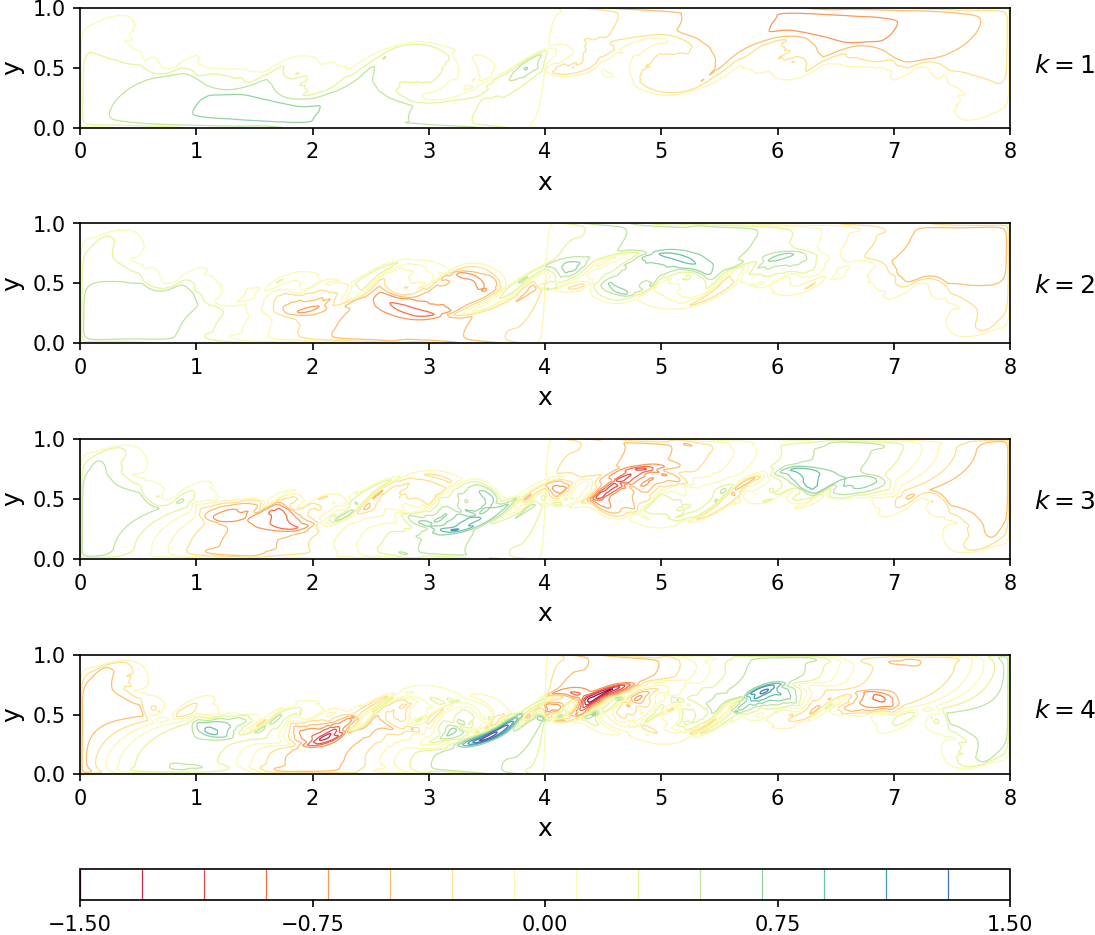}
	\caption{The first 4 global basis functions for temperature field from POD application over the whole time domain (i.e., for $0\le t \le8$) for Boussinesq problem.}
	\label{fig:BQ_basis0}
\end{figure}

\begin{figure}[!ht]
	\centering
	\includegraphics[width=0.45\textwidth]{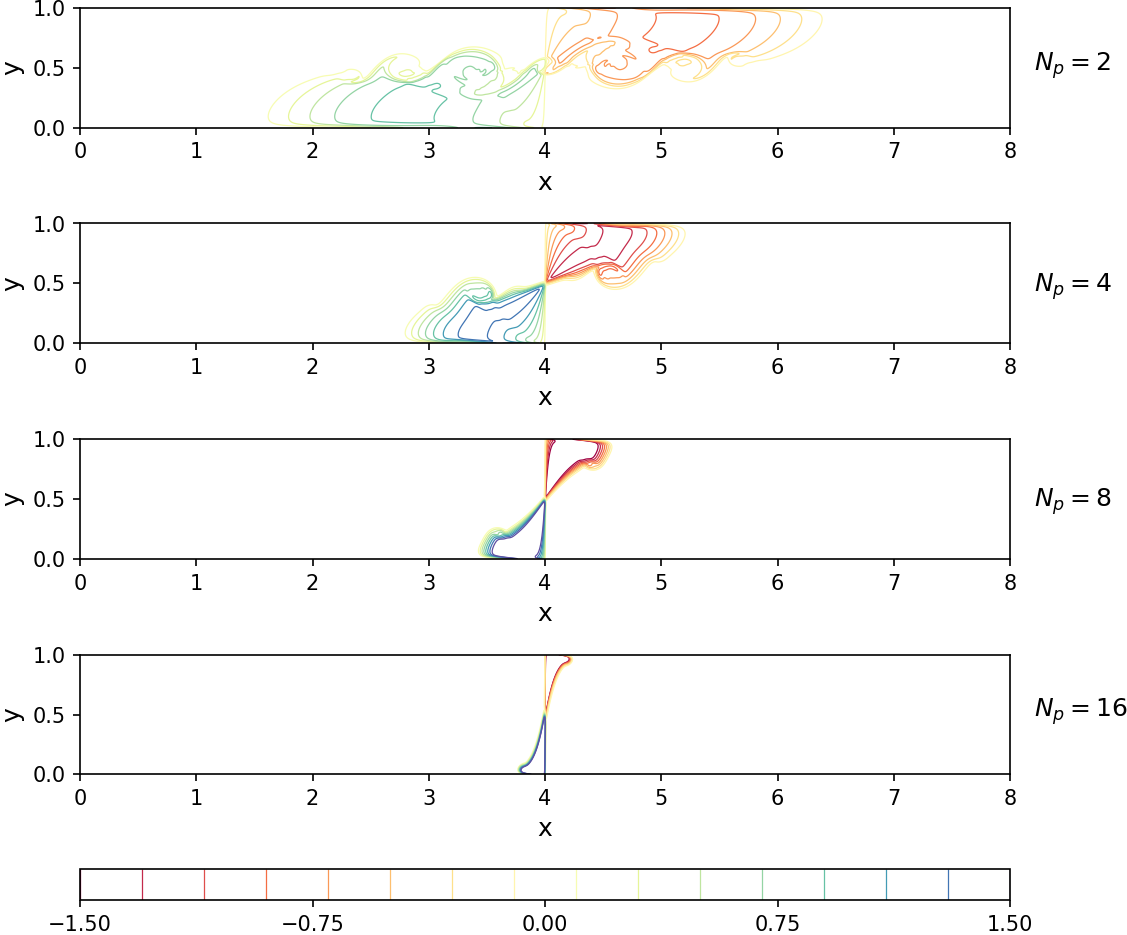}
	\caption{The first local basis function for temperature field from PID application over the first subinterval (i.e., for $0\le t \le t_{\kappa^{(1)}}$) for Boussinesq problem using different number of intervals.}
	\label{fig:BQ_basis1}
\end{figure}

\begin{figure}[!ht]
	\centering
	\includegraphics[width=0.45\textwidth]{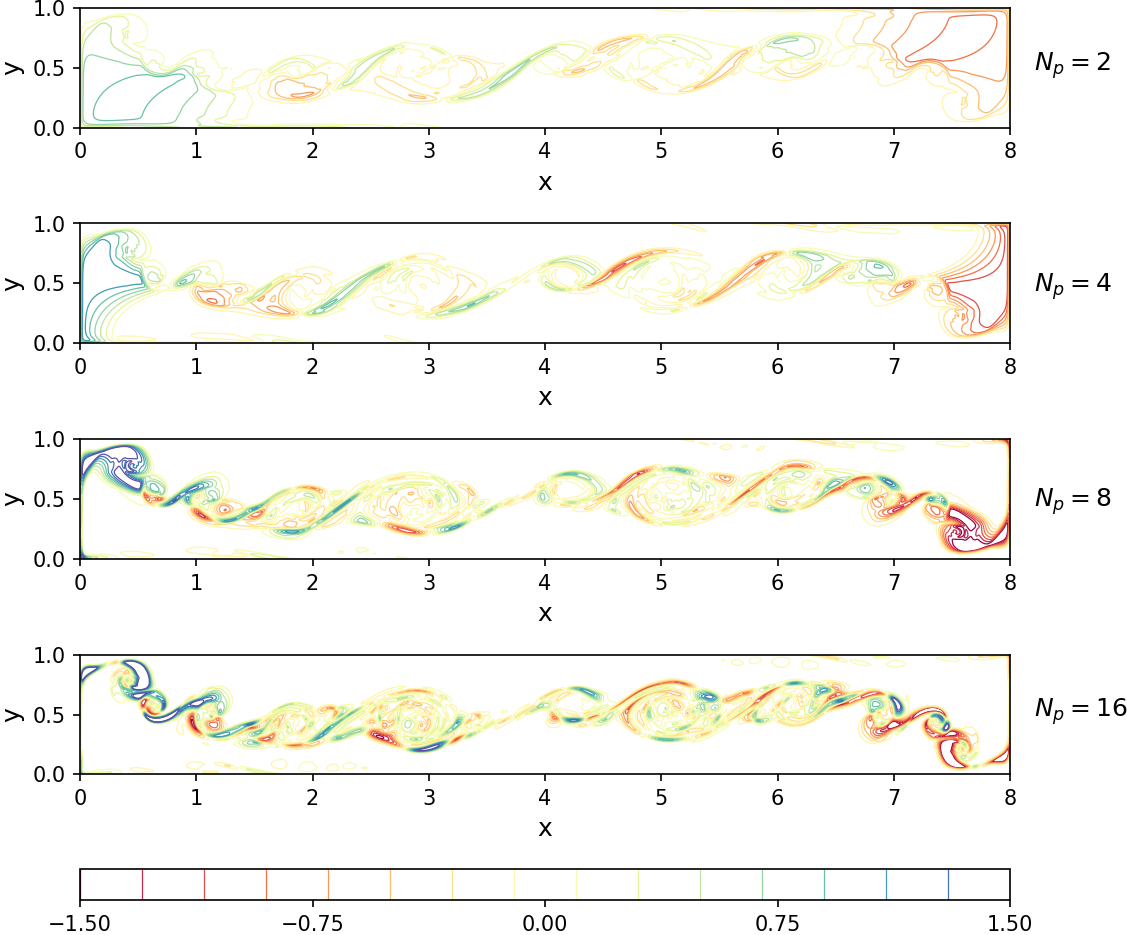}
	\caption{The first local basis function for temperature field from PID application over the first subinterval (i.e., for $t_{\kappa^{(N_p-1)}}\le t \le 8 $) for Boussinesq problem using different number of intervals.}
	\label{fig:BQ_basis2}
\end{figure}

In Boussinesq problem, we provide the first four POD modes for temperature fields in Figure~\ref{fig:BQ_basis0} which obviously demonstrates the averaging nature of POD. We can see that none of these modes looks like the initial or the final fields. They just give an average image of the whole process. On the other hand, applying PID allows us to explore more about the details of the dynamical evolution and underlying instabilities. In Figure~\ref{fig:BQ_basis1}, the initial state of the system can be identified by decomposing the domain into multiple local zones and investigating the first interval where the initial field is located. Similarly, information about the final temperature field can be acquired from narrowing the last partition (i.e., by increasing the number of intervals) as shown in Figure~\ref{fig:BQ_basis2}.

\bibliographystyle{unsrt} 
\bibliography{ref}   

\end{document}